\documentclass[sn-mathphys,Numbered]{sn-jnl}

\usepackage{graphicx}%
\usepackage{multirow}%
\usepackage{amsmath,amssymb,amsfonts}%
\usepackage{amsthm}%
\usepackage{mathrsfs}%
\usepackage[title]{appendix}%
\usepackage{xcolor}%
\usepackage{textcomp}%
\usepackage{manyfoot}%
\usepackage{booktabs}%
\usepackage{algorithm}%
\usepackage{algorithmicx}%
\usepackage{algpseudocode}%
\usepackage{listings}%
\usepackage{lineno}

\theoremstyle{thmstyleone}%
\theoremstyle{thmstyletwo}%

\theoremstyle{thmstylethree}%

\begin{document}

\title[AMOC Collapse]{Substantial Risk of 21$^{\mathrm{st}}$ Century AMOC Tipping even under Moderate Climate Change} 


\author*[1]{\fnm{Ren\'e M.} \spfx{van} \sur{Westen}}\email{r.m.vanwesten@uu.nl}

\author[1]{\fnm{Elian Y.P.}  \sur{Vanderborght}}\email{e.y.p.vanderborght@uu.nl}

\author[1]{\fnm{Michael}  \sur{Kliphuis}}\email{m.kliphuis@uu.nl}

\author[1]{\fnm{Henk A.} \sur{Dijkstra}}\email{h.a.dijkstra@uu.nl}

\affil[1]{\orgdiv{Department of Physics}, \orgname{Institute for Marine and Atmospheric research Utrecht, Utrecht University}, \orgaddress{\street{Princetonplein 5}, \city{Utrecht}, \postcode{3584 CC},  \country{the Netherlands}}}


\abstract{{\bf The Atlantic Meridional Overturning Circulation (AMOC) is a key component 
of the climate system and considered to be a tipping element. 
There is still a large uncertainty on the critical global warming level at which the AMOC will 
start to collapse. 
Here we analyse targeted climate model simulations, together with observations, 
reanalysis products and a suite of state-of-the-art climate model results to reassess this 
critical global warming  level. 
We find a critical threshold of +3$^{\circ}$C global mean surface temperature increase
compared to pre-industrial with  a lower bound of +2.2$^{\circ}$C (10\%-Cl). Such 
global mean surface temperature  anomalies are expected to be reached after 2050.
This means that the AMOC is more likely than not ($\geq 50\%$) to tip within the 
21$^{\mathrm{st}}$ century  under a  middle-of-the-road climate change scenario 
and very likely ($\geq 90\%$) to tip under a high emissions scenario. 
The AMOC collapse induced cooling is shown to be offset by the regional warming 
over Northwestern Europe during the 21$^{\mathrm{st}}$ century, but will still induce 
severe impacts on society. }  
}

\keywords{AMOC collapse, Climate change, Climate impacts}



\maketitle

\section*{Main}

The Atlantic Meridional Overturning Circulation (AMOC) has been identified as a tipping element 
in the climate system \cite{Lenton2008,Armstrong2022}.  The AMOC transports heat and salt through 
the global ocean \cite{Johns2011}, and strong reductions  in its strength 
severely impact  climate over large parts of the world \cite{Orihuela2022, vanWesten2023b, 
vanWesten2024a}.  Given these potentially severe global consequences, it is not surprising that the 
present-day AMOC is closely monitored along several arrays   \cite{Garzoli2013, Srokosz2015, 
Lozier2019} in the Atlantic Ocean. 

The 20-year long observational AMOC strength time series from the RAPID-MOCHA array at 26$^{\circ}$N \cite{Srokosz2015} has been complemented by sea surface temperature (SST)-based reconstructions \cite{Ceasar2018,Estella-Perez2020,Ceasar2021,Worthington2021,Michel2023}.  These AMOC reconstructions indicate a gradual 
weakening of the AMOC by a few Sverdrups (1 Sv = $10^6$ m$^3$s$^{-1}$) since 1900 and 
also suggest that the AMOC is approaching its tipping point \cite{Boers2021, 
Ditlevsen2023}.  However, they may not adequately capture  the AMOC behaviour and are prone 
to false positive signals \cite{vanWesten2024a, Smolders2024}.  

AMOC strength projections from the Coupled Model Intercomparison Project (CMIP) phase 6 
(CMIP6)  do not show evidence of a 21$^{\mathrm{st}}$ century AMOC collapse, and their projections 
appear quite insensitive with respect to the  greenhouse gas emission scenario used \cite{Weijer2020}. 
The sixth Intergovernmental Panel on Climate Change (IPCC) report 
states  with \emph{medium confidence} that an AMOC collapse will not occur during 
the 21$^{\mathrm{st}}$ century \cite{Fox-Kemper2021}.  Moreover, an analysis of critical 
global warming levels of tipping elements \cite{Armstrong2022} finds that 
the AMOC is \emph{likely} to tip only above +4$^{\circ}$C global mean surface temperature 
increase with respect to  the pre-industrial level.  This suggests a relatively stable AMOC, even under 
high greenhouse gas emission scenarios such as the Shared Socioeconomic Pathway 
(SSP) 3-7.0.
 
Recently, a quasi-equilibrium hosing simulation \cite{vanWesten2024a} with the Community Earth 
System Model (CESM) showed that the pre-industrial AMOC collapses under the imposed 
freshwater flux forcing. While this simulation was still idealised as no climate change forcing 
was included, it provides the basis for developing efficient diagnostics for an AMOC collapse 
under transient 21$^{\mathrm{st}}$ century forcing. Here, we conduct additional targeted CESM 
simulations of the 21$^{\mathrm{st}}$ century AMOC under climate change  and compare the results with  
those of  24~CMIP6 models, with the aim to reassess  the critical global warming level 
at which the AMOC starts to collapse.

\section*{AMOC Collapse in the Pre-industrial CESM}

\begin{figure}[h!]

\includegraphics[width=1\columnwidth, trim = {0cm 0cm 0cm 0cm}, clip]{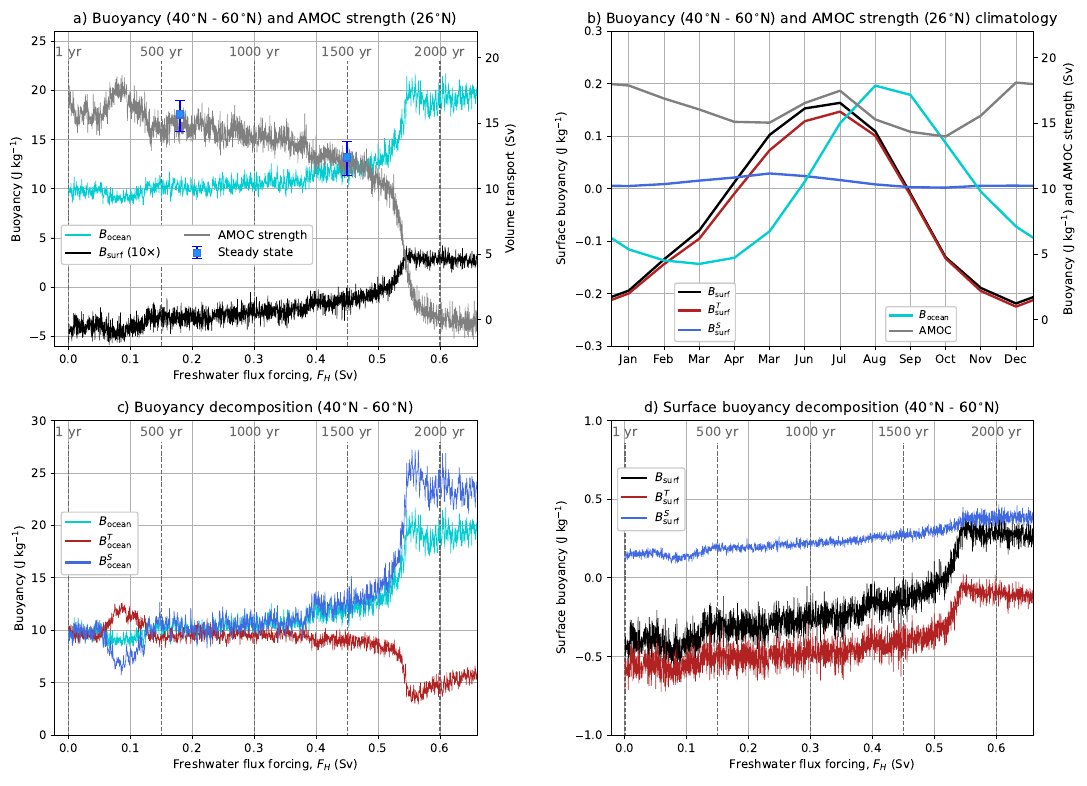}

\caption{\textbf{Buoyancy and AMOC responses under the hosing.}
(a): The AMOC strength at 1,000~m and 26$^{\circ}$N (yearly averages) under the 
quasi-equilibrium  hosing and for the two statistical steady states at constant forcing 
$F_H = 0.18$~Sv and $F_H = 0.45$~Sv. The upper 1,000~m buoyancy ($B_{\mathrm{ocean}}$, 
yearly averages) and surface buoyancy ($B_{\mathrm{surf}}$, yearly sums), spatially 
averaged over the North Atlantic Ocean (40$^{\circ}$N -- 60$^{\circ}$N).
(b): Mean seasonal cycle  over the first 50~model years for the quantities plotted in 
panel~a.
(c): The upper 1,000~m  buoyancy $B_{\mathrm{ocean}}$ (yearly averages), 
decomposition   for the temperature ($B_{\mathrm{ocean}}^T$) and salinity ($B_{\mathrm{ocean}}^S$).
(d): The surface buoyancy $B_{\mathrm{surf}}$ (yearly sums) decomposition for the heat fluxes ($B_{\mathrm{surf}}^T$) and freshwater fluxes ($B_{\mathrm{surf}}^S$).}

\label{fig:Figure_1}
\end{figure}

The quasi-equilibrium hosing simulation (see Methods, \cite{vanWesten2024a}) was conducted 
under constant pre-industrial radiative forcing conditions. The AMOC strength is plotted in 
Figure~\ref{fig:Figure_1}a versus the freshwater flux forcing ($F_H$) which linearly increases from 
0 to 0.66~Sv  over 2,200~model years; the AMOC collapses at about 0.525~Sv (model year~1,758 
\cite{vanWesten2024a}).  From  this simulation the dominant physical  mechanisms of AMOC tipping 
can be quantified and used to analyse  model simulations under climate change. 

Before the collapse (Figure~\ref{fig:Figure_S1}), the AMOC decline is linked to potential density changes over the northern isopycnal outcropping region (located at 40$^{\circ}$N -- 60$^{\circ}$N, Figure~\ref{fig:Figure_S1}c).
The AMOC over the isopycnal outcropping region is in thermal wind balance with the meridional density gradient over this region \cite{Nikurashin2012, Wolfe2014}.
The upper 1,000~m potential density changes over the isopycnal outcropping region primarily drive the AMOC decline and we 
quantify these changes by determining the upper ocean buoyancy ($B_{\mathrm{ocean}}$, see Methods) over the region.
The quantity $B_{\mathrm{ocean}}$ has indeed the opposite response to that of the AMOC strength under the quasi-equilibrium hosing (Figure~\ref{fig:Figure_1}a).

The outcropping of isopycnals at 40$^{\circ}$N -- 60$^{\circ}$N is governed by surface winds, interior mixing and loss of 
surface buoyancy ($B_{\mathrm{surf}}$)  \cite{Nikurashin2012}. Interior ocean isopycnals only surface when $B_{\mathrm{surf}}$ remains net negative over time scales longer than one year. 
There is a strong seasonal cycle in both $B_{\mathrm{ocean}}$ and $B_{\mathrm{surf}}$, where the surface buoyancy leads by about 2~months (Figure~\ref{fig:Figure_1}b).
The surface buoyancy is primarily driven by atmosphere-ocean heat fluxes and the relatively large negative winter values cause strong vertical mixing (i.e., deep convection) over the isopycnal outcropping region.
Once $B_{\mathrm{surf}}$ switches sign, there is no outcropping of isopycnals and only a weak and diffusive AMOC can exist \cite{Wolfe2015}.
The quantity $B_{\mathrm{surf}}$ is linked to water mass transformation and hence AMOC strength \cite{Walin1982, Marshall1999}
and indeed $B_{\mathrm{surf}}$ and AMOC strength have the opposite response under the hosing (Figure~\ref{fig:Figure_1}a).

From the temperature and salinity buoyancy decompositions of $B_{\mathrm{ocean}}$ 
($B_{\mathrm{ocean}}^T$ and $B_{\mathrm{ocean}}^S$, respectively, see Methods) 
it is found that  salinity perturbations weaken the AMOC under the hosing (Figure~\ref{fig:Figure_1}c).   
The non-linear increase in $B_{\mathrm{ocean}}^S$
after 1350~years ($F_H > 0.4$~Sv) indicates a positive salt-advection feedback. 
This feedback is destabilising the AMOC as salinity anomalies are amplified 
through their effect on the AMOC strength and pattern \cite{Marotzke2000, Peltier2014}.
The AMOC weakening and associated reduced meridional heat transport result in lower 
SSTs (i.e., the characteristic AMOC fingerprint \cite{Ceasar2018})  and induce ocean 
buoyancy loss which  stabilises the AMOC. 
The $B_{\mathrm{ocean}}^T$ responses have a passive and opposite role to the $B_{\mathrm{ocean}}^S$ responses. 

The quantity $B_{\mathrm{surf}}$ is decomposed in a similar way (Figure~\ref{fig:Figure_1}d)  
and the $B_{\mathrm{surf}}^S$ increase is primarily driven by the freshwater flux forcing.
Changes in AMOC strength, driven by freshening of the isopycnal outcropping region, reduce the surface ocean and  atmosphere temperatures.
These ocean-atmosphere temperature differences modify the heat  fluxes (Figure~\ref{fig:Figure_S2}) and
show that lower SSTs effectively lose less heat to the atmosphere and reduce the magnitude of $B_{\mathrm{surf}}^T$.
It should be noted that the lower temperatures enhance sea-ice formation conditions around Greenland ($> 60^{\circ}$N) and, via advection of sea ice, increase the sea-ice area over the isopycnal outcropping region.
The expanding sea-ice pack enhances the albedo and  strongly reduces the ocean-atmosphere fluxes  \cite{Lin2023}.
The bottom line is that the freshwater flux forcing increases $B_{\mathrm{surf}}^S$ and, through complex AMOC feedbacks, leads to an equally important contribution in $B_{\mathrm{surf}}^T$.
For example,  $B_{\mathrm{surf}}$ increased by 0.28~J~kg$^{-1}$ for $F_H = 0.45$~Sv 
(model years~1476 -- 1525) compared to beginning of the simulation,
where $B_{\mathrm{surf}}^S$ contributed 47\% and $B_{\mathrm{surf}}^T$  contributed 53\%. 

The isopycnal outcropping region changes from loosing surface buoyancy to gaining surface buoyancy around the AMOC tipping event ($F_H = 0.525$~Sv, model year~1750).
Thereafter the region is gaining net buoyancy through its surface. 
The collapse is caused by the destabilising salt-advection feedback and a positive $B_{\mathrm{surf}}$ ensures that a collapsed state will be reached. 
What is, however, most important from these results is that the buoyancy diagnostics considered can be used to analyse AMOC behaviour in model 
simulations under climate change forcing. 

\section*{The AMOC Collapse under Climate Change}

From the quasi-equilibrium hosing we branched off 500-year long simulations 
under constant freshwater flux forcing  to find statistical equilibria for $F_H = 0.18$~Sv and 
$F_H = 0.45$~Sv (Figure~\ref{fig:Figure_1}a, with more details in \cite{vanWesten2024c}).
The simulation with the higher freshwater flux forcing is closer to the AMOC tipping point under 
constant pre-industrial radiative forcing conditions.
From the end of the statistical equilibria, we branched off the historical forcing (1850 -- 2005) 
followed by the Representative Concentration Pathway (RCP) 4.5 and 8.5 scenarios (2006 -- 2100)
while keeping the value of $F_H$ fixed for each simulation. 
Finally, after the year 2100, we kept the forcing conditions fixed at their 2100~forcing levels 
and ran the simulations for an additional 400~years (up to 2500) to find the statistical equilibria 
under these future forcing conditions.  Ice sheet dynamics are important on such long timescales,
but ice sheets are prescribed in the CESM simulations. Hence the long-term results  only represent 
equilibria under a substantially changed radiative forcing, but with present-day ice sheets. 

\begin{figure}[h!]

\includegraphics[width=1\columnwidth, trim = {0cm 0cm 0cm 0cm}, clip]{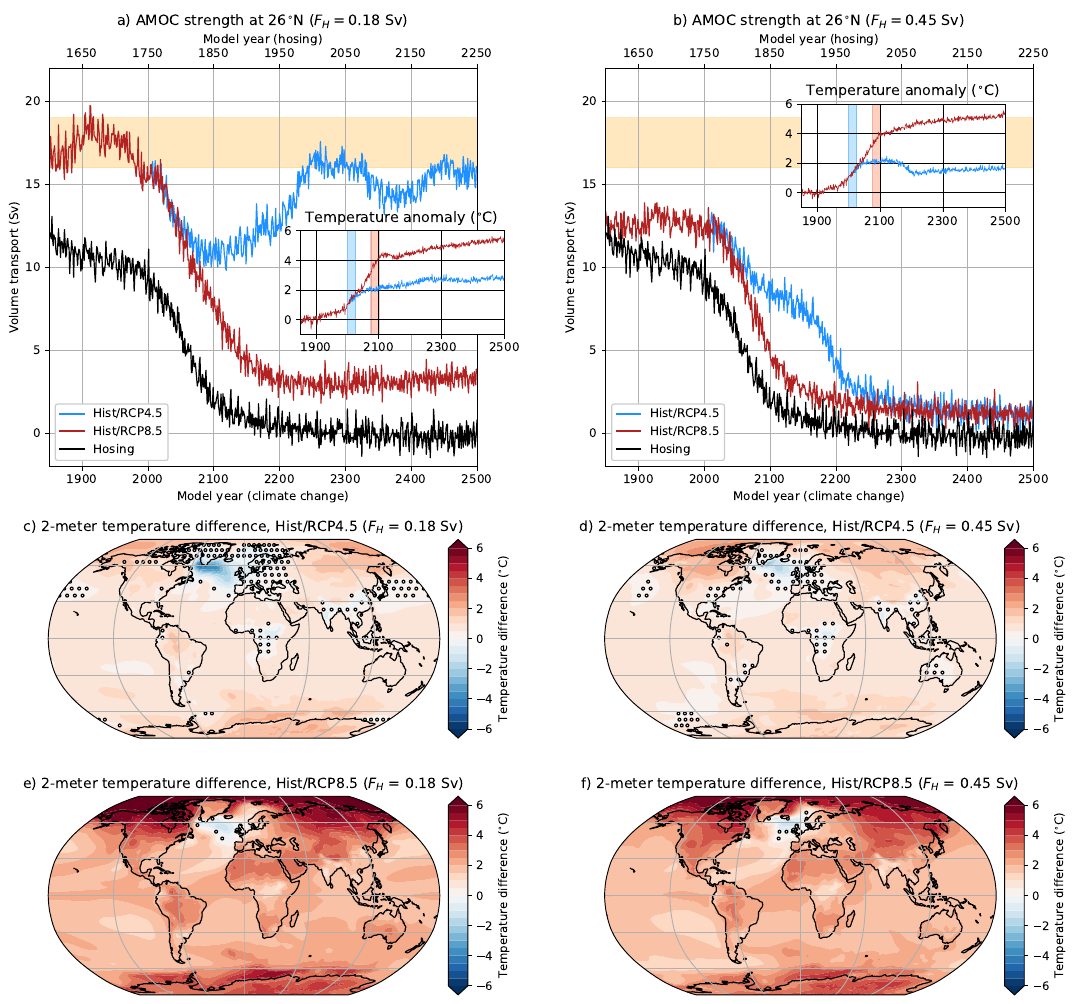}

\caption{\textbf{The AMOC responses under climate change.}
The AMOC strength at 1,000~m and 26$^{\circ}$N under the two climate change scenarios (Hist/RCP4.5 and Hist/RCP8.5) and for (a): $F_H = 0.18$~Sv and (b): $F_H = 0.45$~Sv.
For reference, the AMOC collapse under the hosing is also shown and is centred around the AMOC tipping event (black curve, top x-axis).  
The yellow shading indicates the observed AMOC strength \cite{Smeed2018, Worthington2021}.
The insets show the 2-meter global mean surface temperature anomaly compared to the pre-industrial period (1850 -- 1899).
The yearly-averaged surface temperature differences are determined between the red (2075 -- 2100) and blue (2000 -- 2025) shaded periods
and are presented in panels~c -- f.
The markers in panels~c -- f indicate non-significant ($p \geq 0.05$, two-sided Welch's t-test) differences.}

\label{fig:Figure_2}
\end{figure}

The AMOC strengths at 1,000~m and 26$^{\circ}$N under the two climate change scenarios 
are shown for the $F_H = 0.18$ and  $F_H = 0.45$  cases in Figures~\ref{fig:Figure_2}a,b, 
including the global mean surface temperature anomaly (see insets). For $F_H = 0.18$~Sv, 
the AMOC has a reasonable strength over the historical period compared to present-day observations 
(Figures~\ref{fig:Figure_2}a) but its  freshwater transport at 34$^{\circ}$S (indicated by $F_{\mathrm{ovS}}$) is biased positive  (Figure~\ref{fig:Figure_S3}). 
On the other hand, for  $F_H = 0.45$~Sv,  the freshwater 
transport at 34$^{\circ}$S is in the range of observations (Figure~\ref{fig:Figure_S3}) but the 
historical AMOC strength is weaker than in observations (Figures~\ref{fig:Figure_2}b).  
The quantity $F_{\mathrm{ovS}}$ is an important indicator 
in AMOC tipping dynamics \cite{Dijkstra2007, Huisman2010, Mecking2017, Weijer2019, vanWesten2023b}
and a negative $F_{\mathrm{ovS}}$ sign indicates that the AMOC is closer to its tipping point under the quasi-equilibrium hosing \cite{vanWesten2024a}.

Up to 2020, the AMOC strength remains remarkably constant for the simulations under 
climate change. Thereafter, the AMOC strength substantially weakens by 5 -- 6 Sv under 
Hist/RCP4.5 and 8 -- 9 Sv under Hist/RCP8.5 over the 21$^{\mathrm{st}}$ century.
The AMOC weakening rates under Hist/RCP8.5 are comparable to that of the hosing with 
a decline of 8~Sv over about 100 years (Figures~\ref{fig:Figure_1}a). 
The RCP4.5 and RCP8.5 scenarios show little differences in their AMOC responses up 
to 2050 and this is consistent with the results for the CMIP6  models \cite{Weijer2020}. 

The climate change scenarios for  $F_H = 0.45$~Sv both show AMOC collapses and 
their AMOC strengths reduce to 1~Sv at the end of the simulations.
The AMOC strength in RCP4.5 ($F_H = 0.45$~Sv) reduces by 3.9~Sv (-31\%) in 
2090 -- 2100 (with respect to 1850 -- 1899) and eventually collapses after 2150.
This suggests that,  under this scenario, an AMOC tipping event occurs  somewhere in the 
21$^{\mathrm{st}}$ century. For $F_H = 0.18$~Sv, only the 
AMOC under RCP8.5 collapses and appears to stabilise around 3.5~Sv at the 
end of the simulation. In the RCP4.5 (and $F_H = 0.18$~Sv) scenario the AMOC 
weakens by 6.1~Sv (-36\%) in 2090 -- 2100 (with respect to 1850 -- 1899) and 
eventually recovers to 15.7~Sv (-6.7\%) at the end of the simulation. This shows
that quantifying an AMOC  collapse based on absolute AMOC anomalies or 
fractional changes \citep{Jackson2022b} may misinterpret an imminent 
collapse.  

The extended simulations provide insight into the equilibria resulting under fixed 
future climate forcing. Under RCP8.5 and for $F_H = 0.18$~Sv, the AMOC reduces 
to a weak and shallow ($<$~1000~m)  overturning cell and the quantity $F_{\mathrm{ovS}}$ becomes negative (Figure~\ref{fig:Figure_S3}). 
This is a typical characteristic of an `AMOC weak' state which is an additional statistical equilibrium in 
the full hysteresis hosing simulation under pre-industrial conditions \cite{vanWesten2023b, 
vanWesten2024c}. The AMOC states resulting at  $F_H = 0.45$~Sv have no clear 
overturning cell and their $F_{\mathrm{ovS}}$ values become positive. These 
AMOC characteristics link to an `AMOC off' state as found under pre-industrial conditions 
as a statistical equilibrium \cite{vanWesten2024c}. 

When comparing the two RCP4.5 simulations, the temperature responses over the 
21$^{\mathrm{st}}$ century (Figures~\ref{fig:Figure_2}c,d) are, however, very similar. 
The same holds for the two RCP8.5 simulations over the  21$^{\mathrm{st}}$ century 
(Figures~\ref{fig:Figure_2}e,f).  The AMOC weakening can (partly) offset the regional 
warming over Northwestern  Europe resulting in below-averaged surface temperature 
rise for a few European  cities in the 21$^{\mathrm{st}}$ century (Figures~\ref{fig:Figure_S4}a--d).  
The surface  temperature anomalies at the end of the simulation (i.e., the equilibrium 
temperatures,  model years~2450 -- 2499, Figure~\ref{fig:Figure_S4}e--h) 
clearly show the strong AMOC-induced cooling when the AMOC has collapsed 
and are most pronounced for $F_H = 0.45$~Sv and the RCP4.5 scenario. 

\section*{The Buoyancy Responses under Climate Change}

Also under climate change,  the $B_{\mathrm{ocean}}$ over the isopycnal outcropping region
is  related to the AMOC strength (Figures~\ref{fig:Figure_S5}a--d). 
The $B_{\mathrm{surf}}$ is significantly ($p < 0.05$, \cite{Santer2000}) increasing during the 21$^{\mathrm{st}}$ 
century and the changes are smaller in the Hist/RCP4.5 (0.29~J~kg$^{-1}$ and 0.26~J~kg$^{-1}$)
than in the Hist/RCP8.5 (0.49~J~kg$^{-1}$ and 0.44~J~kg$^{-1}$, for $F_H = 0.18$~Sv and $F_H = 0.45$~Sv, 
respectively). As expected, the larger increase in $B_{\mathrm{surf}}$ under Hist/RCP8.5 (compared to Hist/RCP4.5) 
can be attributed to a larger radiative forcing (Figures~\ref{fig:Figure_S5}a--d).
The ocean buoyancy decomposition into its temperature and salinity contributions 
(Figures~\ref{fig:Figure_S5}e--l) clearly shows that temperature responses are forcing the AMOC.
Prior to the AMOC decline of model year~2020, the temperature is weakening the AMOC by buoyancy gain ($B_{\mathrm{ocean}}^T$) and thereafter salinity responses ($B_{\mathrm{ocean}}^S$) start to dominate.
This strong increase in $B_{\mathrm{ocean}}^S$ indicates a destabilising salt-advection feedback.

The buoyancy responses under climate change are clearly different from those of the quasi-equilibrium 
hosing simulation, because in the latter case  the ocean salinity responses ($B_{\mathrm{ocean}}^S$) are  
entirely responsible for destabilising the AMOC. Another difference is that under climate change the surface 
buoyancy increase is  induced by heat flux ($B_{\mathrm{surf}}^T$) changes and the freshwater flux 
($B_{\mathrm{surf}}^S$) changes have a negligible contribution. However, once the AMOC starts to decline, 
we find comparable (with respect to  the quasi-equilibrium case) responses in the latent, sensible, longwave, 
evaporation and precipitation contributions 
(Figure~\ref{fig:Figure_S6}).   The AMOC sea-ice feedbacks are only relevant for the Hist/RCP4.5 and the 
$F_H = 0.45$~Sv case as the sea ice vastly extends under its AMOC collapse. For the other simulations, 
 the higher background temperatures melt almost all sea ice and hence are limiting the destabilising sea-ice 
 buoyancy contribution. 

The AMOC collapses under the two RCP8.5 scenarios and the dynamics driving the collapse shares similarities 
to that of the hosing simulation (Figure~\ref{fig:Figure_S5}).
The $B_{\mathrm{surf}}$ increase levels off when the climate forcing becomes constant in both RCP4.5 simulations.
Depending on the freshwater flux forcing, this results in either a negative $B_{\mathrm{surf}}$ or positive $B_{\mathrm{surf}}$ in 2100
for $F_H = 0.18$~Sv and $F_H = 0.45$~Sv, respectively.
The AMOC eventually recovers for $F_H = 0.18$~Sv (Figure~\ref{fig:Figure_2}a),
but for $F_H = 0.45$~Sv the isopycnal outcropping region slowly gains buoyancy  between model years~2080 to 2150.
This  results in an AMOC decline of only 1~Sv between 2100 and 2150 and thereafter it fully collapses.
During the AMOC collapse,   $B_{\mathrm{surf}}$ further increases which  is similar to the hosing and RCP8.5 simulations.
A net negative surface buoyancy is crucial for AMOC stability \cite{Walin1982, Marshall1999, Nikurashin2012, Wolfe2014, Wolfe2015} and an 
AMOC collapse is inevitable when $B_{\mathrm{surf}}$ becomes  positive. 

\section*{Surface Buoyancy Responses in  CMIP6 models} 

As the $B_{\mathrm{surf}}$ over the isopycnal outcropping region is a  diagnostic of an imminent AMOC collapse, 
we have determined it  for 24~different CMIP6 models under the historical forcing 
(1850 -- 2014) followed by the SSP2-4.5 and SSP5-8.5 (2015 -- 2100) scenarios. In CMIP6 there are no 
indications  for an AMOC collapse by the year 2100 \cite{Weijer2020, vanWesten2024b}.
This is not very surprising as the typical AMOC collapse timescale is longer than 100~years and 
only (substantial) AMOC weakening can be expected by 2100 (Figures~\ref{fig:Figure_2}a,b). 

\begin{figure}[h!]

\vspace{-0.0cm}
\includegraphics[width=1\columnwidth, trim = {0cm 0cm 0cm 0cm}, clip]{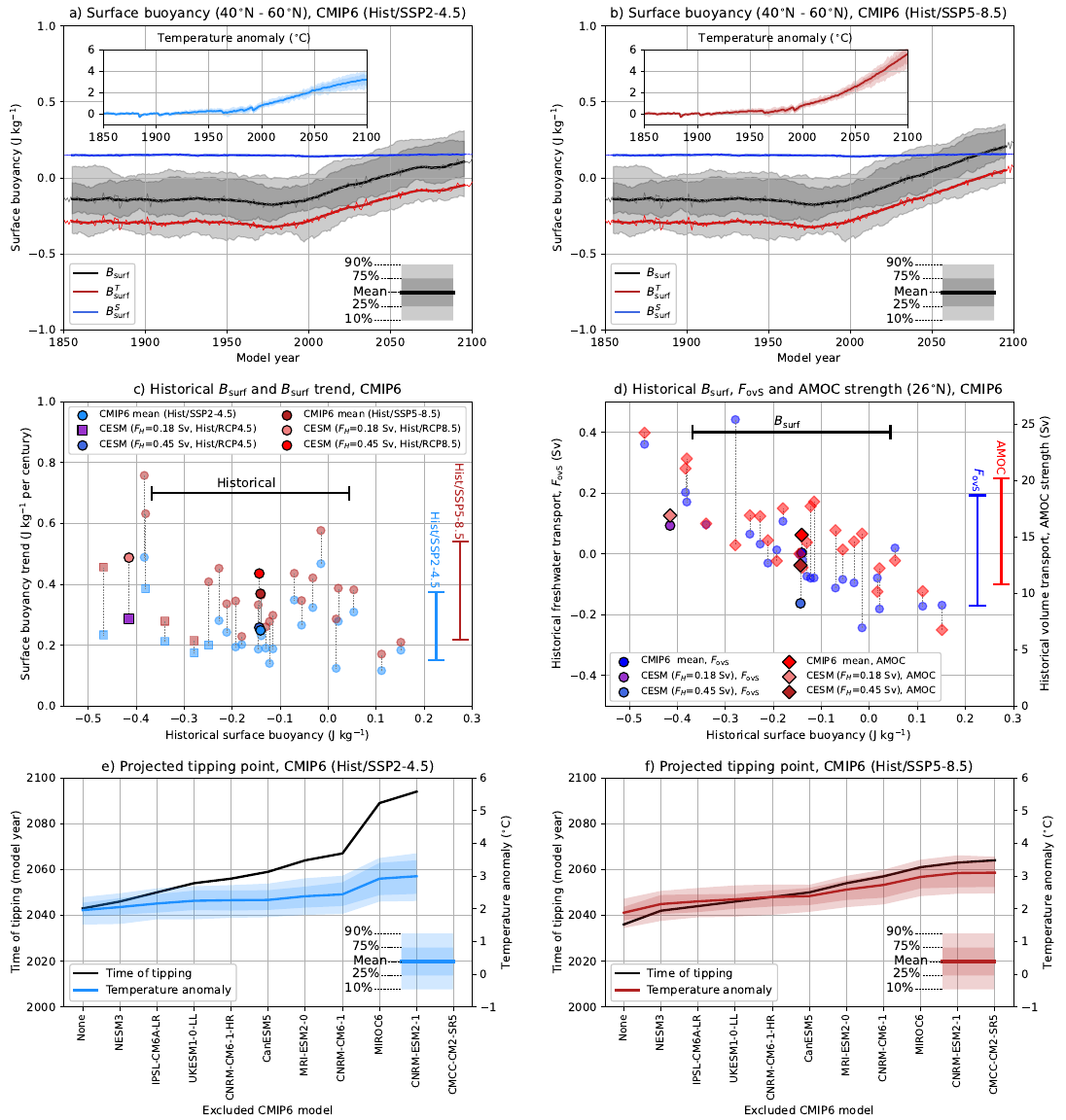}

\caption{\textbf{Surface buoyancy responses  in the CMIP6.}
(a \& b): The surface buoyancy ($B_{\mathrm{surf}}$) decomposition for CMIP6 
under the Hist/SSP2-4.5 and Hist/SSP5-8.5 scenarios. 
The multi-model mean and percentiles are smoothed through a 11-year running mean.
(c): The 24~CMIP6 models with the historical (1850 -- 1899) $B_{\mathrm{surf}}$ against the $B_{\mathrm{surf}}$ trends (2000 -- 2100) for Hist/SSP2-4.5 (blue) and Hist/SSP5-8.5 (red).
The error bars indicate the 10\% to 90\%-Cl. 
The thin lines connect the two forcing scenarios to the same climate model.
The four CESM simulations (Figure~\ref{fig:Figure_2}) are also displayed.
The circled markers indicate at least one 11-year window with positive $B_{\mathrm{surf}}$,
the squared markers indicate that $B_{\mathrm{surf}}$ remains negative.
(d): Similar to panel~c, but now for the historical $B_{\mathrm{surf}}$ against the $F_{\mathrm{ovS}}$ (circled markers) and AMOC strength (diamond markers).
(e \& f): The tipping time is determined where the $B_{\mathrm{surf}}$ multi-model mean switches sign, 
using the smoothed time series from panels~a and b. The same procedure is repeated by excluding 
the NESM3 (highest historical $B_{\mathrm{surf}}$) from the multi-model mean, then the IPSL-CM6A-LR 
and NESM3 are excluded, and so on. The associated temperature anomaly (over the same 11-year 
window) is also shown.} 

\label{fig:Figure_3}
\end{figure}

The behaviour of $B_{\mathrm{surf}}$ in CMIP6 models is similar  to that in the CESM simulations 
discussed above and the heat fluxes  primarily drive the increase over the 21$^{\mathrm{st}}$ century (Figures~\ref{fig:Figure_3}a,b). 
The CMIP6 multi-model mean $B_{\mathrm{surf}}$ trends (2000 -- 2100) are 0.25 (0.15 -- 0.37, 10\% -- 90\%-Cl)~J~kg$^{-1}$ per century and 
0.37 (0.22 -- 0.54, 10\% -- 90\%-Cl)~J~kg$^{-1}$ per century for Hist/SSP2-4.5 and Hist/SSP5-8.5, respectively (Figure~\ref{fig:Figure_3}c and Table~\ref{tab:Table_S1}).
The multi-model mean $B_{\mathrm{surf}}$ switches sign by 2043 and by 2036 with an associated global warming of
+1.96$^{\circ}$C (+1.52$^{\circ}$C to +2.37$^{\circ}$C, 10\% -- 90\%-Cl, years~2038 -- 2048) and 
+1.88$^{\circ}$C (+1.41$^{\circ}$C to +2.31$^{\circ}$C, 10\% -- 90\%-Cl, years~2031 -- 2041) for Hist/SSP2-4.5 and Hist/SSP5-8.5, respectively (Figures~\ref{fig:Figure_3}e,f).

The magnitude of the $B_{\mathrm{surf}}$ trends is not dependent on the historical $B_{\mathrm{surf}}$ value
and larger $B_{\mathrm{surf}}$ trends can be expected for higher emission scenarios (Figure~\ref{fig:Figure_3}c).
It should be noted that five~CMIP6 models have a positive historical $B_{\mathrm{surf}}$ (1850 -- 1899, see Table~\ref{tab:Table_S1}).
For example, the NESM3 (highest historical $B_{\mathrm{surf}}$) has a very weak historical AMOC strength of 6.8~Sv (Figure~\ref{fig:Figure_3}d)
and, in combination with its negative historical $F_{\mathrm{ovS}}$, suggest that the NESM3 reached an `AMOC weak' state upon the historical initialisation.
On the other hand, the FGOALS-g3 (lowest historical $B_{\mathrm{surf}}$) has a strong historical AMOC strength of 24.3~Sv and
a positive $F_{\mathrm{ovS}}$ of 0.36~Sv and its $B_{\mathrm{surf}}$ does not change sign.

There is a tendency among the suite of CMIP6 models where models with relatively large $B_{\mathrm{surf}}$ have relatively low $F_{\mathrm{ovS}}$ values ($R^2 =  0.66$)
and weak AMOC strengths ($R^2 =  0.69$) for the historical period.
Similar relations in CMIP6 are found for the present-day (1994 -- 2020) period and CMIP6 biases in $F_{\mathrm{ovS}}$ and AMOC strength influence their responses under climate change \cite{vanWesten2024b}.
Previous work \cite{Liu2017} and the CESM results here (Figures~\ref{fig:Figure_2} and \ref{fig:Figure_S3}) show that the case with historical positive $F_{\mathrm{ovS}}$ and strong AMOC is more resilient under climate change
than the case with historical negative $F_{\mathrm{ovS}}$ and weak AMOC strength.
This indicates that the various model biases \cite{vanWesten2024b} upon initialisation set the AMOC sensitivity under the applied forcing.

Excluding the five~CMIP6 models with positive historical $B_{\mathrm{surf}}$ from the multi-model mean
shifts the $B_{\mathrm{surf}}$ sign change to 2059 (warming of +2.27$^{\circ}$C) and 2050 (warming of +2.39$^{\circ}$C) for Hist/SSP2-4.5 and Hist/SSP5-8.5, respectively  (Figures~\ref{fig:Figure_3}e,f).
For the 19~CMIP6 models with a negative historical $B_{\mathrm{surf}}$, we report that 8 (10) models have switched $B_{\mathrm{surf}}$ sign by mid 21$^{\mathrm{st}}$ century and
14 (16)~models by the end of the 21$^{\mathrm{st}}$ century under Hist/SSP2-4.5 (Hist/SSP5-8.5).
Based on the $B_{\mathrm{surf}}$ responses, we expect that most CMIP6 models will show an AMOC tipping event when integrating the two SSP scenarios beyond 2100.

\section*{The Present-day Forcing on the AMOC}

\begin{figure}[h!]

\includegraphics[width=1\columnwidth, trim = {0cm 0cm 0cm 0cm}, clip]{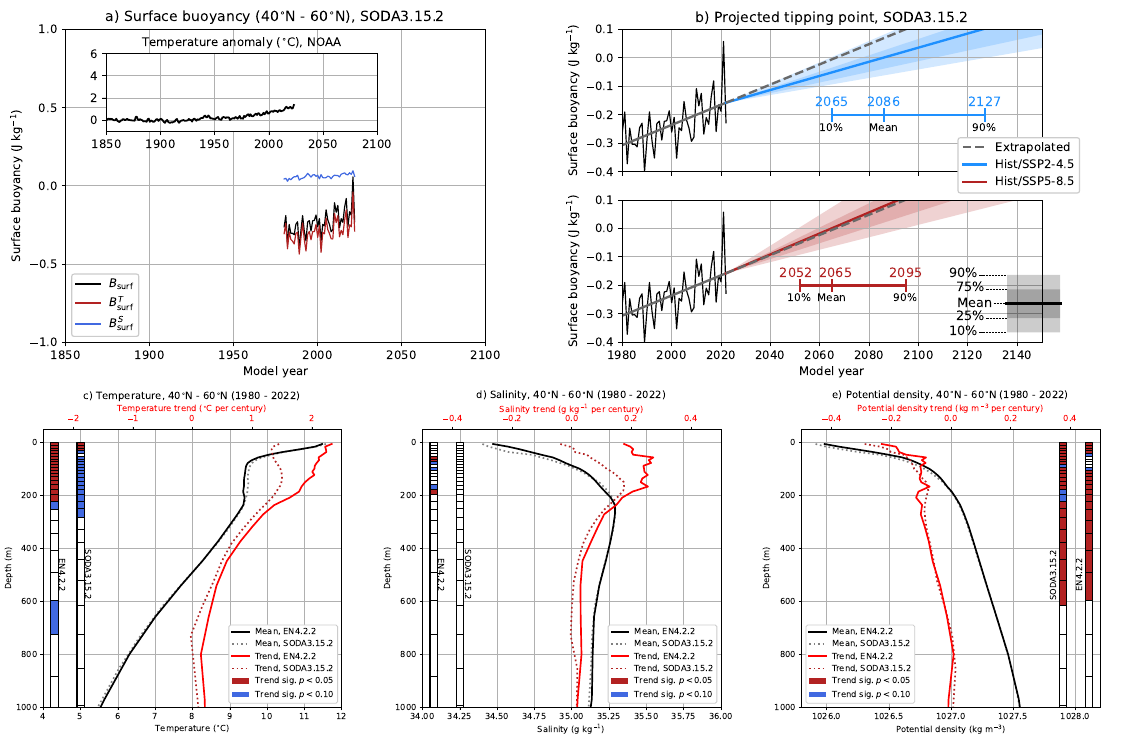}

\caption{\textbf{The tipping point estimate based on surface buoyancy changes under climate change.}
(a): The surface buoyancy ($B_{\mathrm{surf}}$) decomposition for SODA3.15.2 (1980 -- 2022) and 
the observed yearly-averaged global mean surface temperature anomaly compared to the historical period (1850 -- 1899). 
(b): The linearly extrapolated $B_{\mathrm{surf}}$ for SODA3.15.2 (dashed gray curve) and the CMIP6 multi-model mean $B_{\mathrm{surf}}$ trend (over the 21$^{\mathrm{st}}$ century) 
for the Hist/SSP2-4.5 and Hist/SSP5-8.5 scenarios.
(c -- e): The vertical structure of temperature, salinity and potential density over 40$^{\circ}$N -- 60$^{\circ}$N (Atlantic Ocean) and for the period 1980 -- 2022.
The black (red) curves are the time means (trends) for 1980 -- 2022, and are shown for EN4.2.2 and SODA3.15.2.
The significance \cite{Santer2000} of the trends with depth are displayed on the sides of each panel.}

\label{fig:Figure_4}
\end{figure}

Observational data  is very limited  to determine $B_{\mathrm{surf}}$ responses and compare against CMIP6 under 
climate change. Instead, we use the reanalysis product SODA3.15.2 (1980 -- 2022) to analyse $B_{\mathrm{surf}}$ 
responses under present-day climate change. 
The SODA3.15.2 has a mean AMOC strength of 16.8~Sv (not shown) and its $B_{\mathrm{surf}}$ is negative ($-0.23$~J~kg$^{-1}$) and is increasing  (Figure~\ref{fig:Figure_4}a) under climate change by 0.36~J~kg$^{-1}$ 
per century ($p < 0.05$).  The CMIP6 multi-model mean has, as explained above, a similar trend under the 
Hist/SSP5-8.5 scenario  (Figure~\ref{fig:Figure_3}b). 
When we linearly extrapolate the SODA3.15.2 trend to zero (Figure~\ref{fig:Figure_4}b), we find that $B_{\mathrm{surf}}$ changes sign by 2067 and when
using the CMIP6 models this is by 2086 (2065 -- 2127, 10\% -- 90\%-Cl) under Hist/SSP2-4.5 and by 2065 (2052 -- 2095, 10\% -- 90\%-Cl) under Hist/SSP5-8.5.
The associated global warming in the CMIP6 multi-model mean is +2.99$^{\circ}$C (+2.16$^{\circ}$C to +3.71$^{\circ}$C, 10\% -- 90\%-Cl, years~2081 -- 2091) for SSP2-4.5 
and +3.37$^{\circ}$C (+2.52$^{\circ}$C to +3.93$^{\circ}$C, 10\% -- 90\%-Cl, years~2060 -- 2070) for SSP5-8.5.

Because a positive $B_{\mathrm{surf}}$ is not consistent with the negative present-day $B_{\mathrm{surf}}$ found in SODA3.15.2, 
we exclude the 13~CMIP6 models with a $B_{\mathrm{surf}}$ sign change before 2022 in any of the two Hist/SSPs scenarios (Table~\ref{tab:Table_S1}). 
This results in a $B_{\mathrm{surf}}$ sign change by 2084 (2063 -- 2107, 10\% -- 90\%-Cl, Hist/SSP2-4.5) and by 2060 (2047 -- 2079, 10\% -- 90\%-Cl, Hist/SSP5-8.5).
The associated warming levels in the CMIP6 mean (of 11~models) are now +2.78$^{\circ}$C (+2.12$^{\circ}$C to +3.45$^{\circ}$C, 10\% -- 90\%-Cl, Hist/SSP2-4.5) 
and +2.89$^{\circ}$C (+2.28$^{\circ}$C to +3.40$^{\circ}$C, 10\% -- 90\%-Cl, Hist/SSP5-8.5).
The critical temperature anomaly for AMOC tipping is about 3$^{\circ}$C and is robust for the two different forcing scenarios.
The projected tipping time is found after 2050, but the timing strongly depends on the forcing scenario.
Based on the +3$^{\circ}$C warming threshold derived from the SODA3.15.2 extrapolation, 
none (SSP2-4.5) and only 3 (SSP5-8.5) of the 24~CMIP6 models surpass this threshold by mid century (2045 -- 2055) 
and 15 of the CMIP6 (SSP2-4.5) and all of CMIP6 (SSP5-8.5) surpass it by the end of the century (2090 -- 2100).

The positive $B_{\mathrm{surf}}$ trend in SODA3.15.2 suggests that the AMOC is weakening under present-day climate change.
Another line of evidence of AMOC weakening is the vertical stratification increase over the isopycnal outcropping region (Figures~\ref{fig:Figure_4}c--e), which enhances the upper ocean buoyancy. 
The time means (1980 -- 2022) of SODA3.15.2 are very close to the observational product EN4.2.2.
When comparing the temperature and salinity trends between the two products, there are some deviations and these are mainly found in the upper 200~m.
These differences cancel out for the potential density trends, which  are comparable between the two products and, more importantly, are primarily induced by positive temperature trends.
Previous studies  \cite{Levang2020, Bonan2022} and our buoyancy decomposition of the CESM under climate change (Figure~\ref{fig:Figure_S5}) demonstrate that global warming enhances the vertical temperature stratification
and induces the initial AMOC weakening.

Satellite observations between 1993 and 2020 show the characteristic warming hole south of Greenland (Figure~\ref{fig:Figure_S7}) and this has been
linked to AMOC weakening \cite{Ceasar2018}, but this interpretation has received substantial criticism \cite{Little2020,Worthington2021,He2022}.
The products EN4.2.2 and SODA3.15.2, together with reanalysis products GLORS12V1 and ORAS5, realistically reproduce the observed SST trends over the same period (Figure~\ref{fig:Figure_S7}).
There are, however, differences in the sea surface salinity trends when comparing the different products. 
The time mean temperature, salinity and potential density with depth are consistent among the products, 
whereas the trends are less coherent.
The ORAS5 has the largest trend differences compared to EN4.2.2, in particular for salinity, and its surface buoyancy does not show a significant trend over its total available period (1985 -- 2023, Figure~\ref{fig:Figure_S8}a).
The GLORYS12V1 does not provide the required data to determine the surface buoyancy.
Both observational and reanalysis products are too short (28~years) to reliably test their significance on the trends. 
Nevertheless, the results over the 1993 -- 2020 period also indicate that positive temperature anomalies 
dominate the vertical stratification responses under climate change, leading to an increase of the upper 
ocean buoyancy. 

\section*{Discussion}

In three out of the four climate change CESM simulations shown here we find that the AMOC 
collapses to a much ($<-80\%$)  weaker state. This weaker state is reached after the 21$^{\mathrm{st}}$ 
century but a  substantial AMOC  weakening occurs already during the 21$^{\mathrm{st}}$ century. 
The initial AMOC weakening under climate change is induced by transient temperature responses 
\cite{Levang2020, Bonan2022} and thereafter the dominant salt-advection feedback takes over 
\cite{Gerard2024}. The AMOC-induced cooling over Northwestern Europe is strong enough to offset 
the regional warming under climate change. 

We have linked the AMOC collapse  to the surface buoyancy over the isopycnal outcropping region in the 
Atlantic Ocean (40$^{\circ}$N -- 60$^{\circ}$N) for which there is a strong theoretical support 
\cite{Nikurashin2012, Wolfe2014, Wolfe2015}. 
By first analysing the quasi-equilibrium hosing simulation (without any climate change), 
we find that a diagnostic for the AMOC tipping event is the sign change in the characteristic net surface buoyancy loss. 
The destabilising salt-advection feedback, in combination with surface buoyancy gain, results in an AMOC tipping event in the CESM.
%
Under the high emission scenario (Hist/RCP8.5), the surface buoyancy always changes sign in the 21$^{\mathrm{st}}$ century and the AMOC collapses  beyond 2100.
In the middle-of-the-road scenario (Hist/RCP4.5), the AMOC weakening and surface buoyancy increase are smaller compared to the high emission scenario. In this scenario, depending on the background climate  conditions and thus the initial surface buoyancy value, this either results in (partial) AMOC recovery or in a full AMOC collapse. 

The reanalysis product SODA3.15.2 shows an increase in the surface buoyancy and vertical stratification and both indicate that the AMOC is weakening under present-day climate change.
By linearly extrapolating the CMIP6 projections, the AMOC tipping event occurs by 2065 (Hist/SSP5-8.5) and by 2086 (Hist/SSP2-4.5)
and corresponds to a global warming of about +3$^{\circ}$C (+2.2$^{\circ}$C to +3.9$^{\circ}$C, 10\% -- 90\%-Cl).
More than half of the CMIP6 models under SSP2-4.5 surpasses the +3$^{\circ}$C warming threshold by the end of the century and all of them under SSP5-8.5,
indicating that there is a substantial risk of AMOC tipping under moderate climate change.
The full AMOC collapse develops after 2100 and it is therefore important to simulate to at least 2200 in the next generation climate models of CMIP phase~7. 

The quantity $F_{\mathrm{ovS}}$ is an important diagnostic and a negative $F_{\mathrm{ovS}}$ sign appears to be connected to the salt-advection feedback strength \cite{Dijkstra2007, Huisman2010, Mecking2017, Weijer2019, vanWesten2024a}.
Note that this quantity can only be used under quasi-equilibrium conditions
and these conditions include the equilibration of the lower level interior basin \cite{Bonan2022} in response to an imposed anomaly.
However, if the deep ocean has no time to equilibrate as a result of a large  freshwater flux change \cite{Orihuela2022, 
Jackson2022b} or rapid climate change, the $F_{\mathrm{ovS}}$ sign is a less clear indicator of the salt-advection feedback. 
Until now, the observed AMOC decline is only gradual \cite{Ceasar2018,Estella-Perez2020,Ceasar2021,Worthington2021,Michel2023} and
the negative $F_{\mathrm{ovS}}$ trend \cite{vanWesten2024a} is still a useful early warning indicator. 
Alternatively, the surface buoyancy is linked to the $F_{\mathrm{ovS}}$ and AMOC strength and can be used under 
strong transient conditions as collapse indicator. 
These relations are found for the historical period, the present-day (1994 -- 2020) period \cite{vanWesten2024b} and in a strongly eddying version of the CESM (Figure~\ref{fig:Figure_S8}b).
The bottom line is that climate model biases \cite{Jackson2013, Drijfhout2011, Mecking2016, Liu2017, Mecking2017, vanWesten2024b, Bonan2024, Dijkstra2024b} need to be reduced to have a realistic AMOC sensitivity under climate change.

Recent AMOC studies have (indirectly) questioned the sixth IPCC report's statement \cite{Fox-Kemper2021}
on the possibility of an AMOC collapse before 2100 \cite{Boers2021, vanWesten2024a, vanWesten2024b, Lohmann2024}.
Our study demonstrates again that the present-day AMOC is on route to tipping.
The previous critical temperature threshold of +4$^{\circ}$C (+1.4$^{\circ}$C to +8.0$^{\circ}$C) warming \cite{Armstrong2022}
was based on enhanced ice and river run-off \cite{Lenton2008, Rahmstorf2005} and the results from a few CMIP phase~5 models \cite{Drijfhout2015, Sgubin2017}.
We link AMOC dynamics to the surface buoyancy and, by using reanalysis and 24~different CMIP6 models, 
we arrive at a critical temperature threshold of +3$^{\circ}$C (+2.2$^{\circ}$C to +3.9$^{\circ}$C) warming for causing 
an AMOC tipping event.
Under high emission scenarios this temperature threshold will be reached after 2050 and is in agreement with previous estimates \cite{Ditlevsen2023, Smolders2024}.
Note that these temperature thresholds are upper bounds as enhanced Greenland Ice Sheet melt under future warming is not considered in the CESM nor CMIP6.
A positive note is that a lower emission scenario can delay or prevent an AMOC collapse, as was demonstrated for the CESM
here. 
To limit the risk of a potential AMOC collapse in the foreseeable future, 
global society needs to be on track of a low emission scenario (i.e., SSP1-1.9 and SSP1-2.6) and urgent climate action is needed to guarantee this.

\newpage
\backmatter

\section*{Methods}

\bmhead{Climate Model Simulations}

The CESM is a fully-coupled climate model and has horizontal resolutions of 1$^{\circ}$ for the ocean/sea-ice components
and 2$^{\circ}$ for the atmosphere/land model components.
Mesoscale features such as ocean eddies are parameterised.
We imposed a slowly-varying freshwater flux forcing (i.e., quasi-equilibrium approach) over the North Atlantic Ocean (20$^{\circ}$N --  50$^{\circ}$N)
and compensated elsewhere to conserve the total ocean salinity \cite{vanWesten2024a}.
From the quasi-equilibrium simulation, we branched off two simulations under their 
constant freshwater flux forcing of $F = 0.18$~Sv and $F = 0.45$~Sv to find statistical equilibria \cite{vanWesten2024c}.
From the end of these statistical equilibria simulations, we imposed the historical forcing (1850 -- 2005) followed by either the RCP4.5 or RCP8.5 scenario (2006 -- 2100)
and with a constant freshwater flux forcing ($F = 0.18$~Sv or $F = 0.45$~Sv). \\

The high-resolution version of the CESM was retained from the  iHESP project \cite{Chang2020} under the Hist/RCP8.5 scenario (and no freshwater flux forcing).
The high-resolution CESM version has a 0.1$^{\circ}$ for the ocean/sea-ice components
and 0.25$^{\circ}$ for the atmosphere/land model components, 
and explicitly resolves ocean eddies and tropical cyclones.
A companion low-resolution version of the CESM is available with the iHESP project, however, not all relevant variables were stored to determine the surface buoyancy
and hence we did not include this simulation in our analysis. \\

For the CMIP6 models, we retained the historical forcing (1850 -- 2014) followed by the SSP2-4.5 and SSP5-8.5 (2015 -- 2100) scenarios.
Note that the forcing scenarios are slightly different between the CESM simulations (the RCPs) and CMIP6 (the SSPs).
We retained the sea surface temperature (`tos'), sea surface salinity (`sos'), total heat flux (`hfds'), total freshwater flux (`wfo'), meridional velocity (`vo'), salinity (`so') and surface air temperature (`tas').
Only 24~models (Table~\ref{tab:Table_S1}) provided all seven~variables for the different forcing scenarios and we selected one realisation (`r1i1p1f1' or `r1i1p1f2') for each CMIP6 model,
the analysis is conducted on their native grid. \\

The observational product EN4.2.2 (horizontal resolution of 1$^{\circ}$ $\times$ 1$^{\circ}$, 1900 -- 2023) assimilates observations from various sources \cite{Gouretski2010}.
The reanalysis products are steered towards observations and the products are 
the SODA3.15.2 (horizontal resolution of 1/4$^{\circ}$ $\times$ 1/4$^{\circ}$, 1980 -- 2022),
the GLORYS12V1 (horizontal resolution of 1/12$^{\circ}$ $\times$ 1/12$^{\circ}$, 1993 -- 2020)
and the ORAS5 (horizontal resolution of 1/4$^{\circ}$ $\times$ 1/4$^{\circ}$, 1958 -- 2023).

\bmhead{The AMOC strength} The AMOC strength is defined as the total meridional volume transport at 26$^{\circ}$N over the upper 1,000~m:
 \begin{equation} \label{eq:AMOC}
\mathrm{AMOC}(y = 26^{\circ}\mathrm{N}) = \int_{-1000}^{0} \int_{x_W}^{x_E} v~\mathrm{d}x \mathrm{d}z
 \end{equation}
When thermal wind balance is applicable for the AMOC (AMOC$_{\mathrm{TWB}}$), its strength can be approximated by \cite{Nikurashin2012, Jansen2018}:
 \begin{equation} \label{eq:AMOC_TWB}
\partial_{zz} \Psi =  \frac{g}{\rho_0 f} \left( \sigma_{\mathrm{b}}(z) - \sigma_{\mathrm{n}}(z) \right)
 \end{equation}
with $g = 9.8$~m~s$^{-2}$, $\rho_0 = 1027$~kg~m$^{-3}$, $f = 1.2 \times 10^{-4}$~s$^{-1}$.
The terms $\sigma_{\mathrm{b}}$ and $\sigma_{\mathrm{n}}$ represent the (spatially-averaged) potential density over the
Atlantic basin (34$^{\circ}$S -- 40$^{\circ}$N) and the isopycnal outcropping region (40$^{\circ}$N -- 60$^{\circ}$N) with depth.
The AMOC$_{\mathrm{TWB}}$ represents the AMOC strength over the isopycnal outcropping region ($\approx 50^{\circ}$N)
and captures variability on decadal to centennial time scales.
We determined the AMOC$_{\mathrm{TWB}}$ at 1,000~m depth (Figures~\ref{fig:Figure_S1}e,f).
The potential densities are determined using the Thermodynamic Equation of SeaWater 2010 (TEOS-10) toolkit \cite{Mcdougall2011}, 
which uses ocean temperature and salinity as input. 
Note that the CESM provides the potential density as standard output, 
but for consistency with the buoyancy calculation (see below) and other reanalysis products which do not provide the potential density as standard output, we used the TEOS-10 toolkit.

\bmhead{Oceanic Buoyancy} The ocean buoyancy and surface buoyancy \cite{Bilo2022} are defined as:

\begin{subequations} \label{eq:Buoyancy}
\begin{align}
B_{\mathrm{ocean}}(T, S) &= \frac{g}{\rho_0} \int_{z_1}^{z_2} \left( \sigma(z) - \sigma(z=0) \right) \mathrm{d}z \\
B_{\mathrm{surf}}	      &= \frac{g \alpha}{\rho_0 C_p} Q_{\mathrm{heat}} + \beta g S_{\mathrm{surf}} Q_{\mathrm{fresh}} = B_{\mathrm{surf}}^T + B_{\mathrm{surf}}^S
\end{align}
\end{subequations}
where $\sigma$ is the potential density (in kg~m$^{-3}$), 
$\alpha$ the thermal expansion coefficient (in K$^{-1}$), $C_p$ the oceanic heat capacity (in J~kg$^{-1}$~K$^{-1}$), 
$Q_{\mathrm{heat}}$ the net heat input into the ocean (in J), $\beta$ the haline contraction coefficient (in kg~g$^{-1}$), $S_{\mathrm{surf}}$ the sea surface salinity (in g~kg$^{-1}$) 
and $Q_{\mathrm{fresh}}$ the net freshwater input into the ocean (in m).
The term $Q_{\mathrm{heat}}$ ($Q_{\mathrm{fresh}}$) can be further decomposed in, for example, 
the contribution by longwave radiation and sensible heat fluxes (evaporation and freshwater flux forcing).

The temperature (salinity) contribution to ocean buoyancy changes can be determined by using a reference salinity (temperature) and
 is indicated by $B_\mathrm{ocean}(T, \overline{S}) =  B_{\mathrm{ocean}}^T$ ($B_\mathrm{ocean}(\overline{T}, S) =  B_{\mathrm{ocean}}^S$).
The monthly-averaged references are determined over the first 50~years of the hosing simulation and the historical period (1850 -- 1899).
The sea water properties (e.g., $\alpha$, $C_p$, etc.) are determined using TEOS-10.
The ocean buoyancy and surface buoyancy are determined with monthly-averaged model output and are then converted to yearly averages and yearly sums, respectively. 

\bmhead{The Freshwater Transport}

The freshwater transport carried by the overturning component  at 34$^{\circ}$S ($F_{\mathrm{ovS}}$) is determined as: 
\begin{equation}
F_{\mathrm{ovS}}= F_{\mathrm{ov}}(y = 34^{\circ}\mathrm{S}) = - \frac{1}{S_0} \int_{-H}^{0} \left[ \int_{x_W}^{x_E} v^* \mathrm{d} x \right] \left[ \langle S \rangle - S_0 \right] \mathrm{d}z
\end{equation}
where $S_0 = 35$~g~kg$^{-1}$ is a reference salinity. The $v^*$ is defined as $v^* = v - \hat{v}$,
where $v$ is the meridional velocity and $\hat{v}$  the full-depth section spatially-averaged meridional velocity.
The quantity $\langle S \rangle$ indicate the zonally-averaged salinity and primed quantities ($v'$ 
and $S'$) are deviations from their respective zonal means.

\bmhead{Acknowledgments}

The model simulation and the analysis of all the model output was conducted on the Dutch National 
Supercomputer Snellius within NWO-SURF project 17239. 

\section*{Declarations}

\begin{itemize}
\item Funding -- R.M.v.W., E.Y.P.V. and H.A.D. are funded by the European Research Council through the ERC-AdG project TAOC (project 101055096). 
\item Conflict of interest -- The authors declare no competing interest
\item Ethics approval -- Not applicable
\item Availability of data and materials -- The (processed) model output will be made available on Zenodo upon publication.
The reanalysis and assimilation products can be accessed through: SODA3.15.2 (http://www.soda.umd.edu), EN4.2.2.g10 (https://www.metoffice.gov.uk/hadobs/en4/),
GLORYS12V1 (https://doi.org/10.48670/moi-00021) and ORAS5 (https://doi.org/10.24381/cds.67e8eeb7).
The observational products are found through:
sea surface temperature observations (version~2.1, Level~4, https://doi.org/10.24381/cds.cf608234), 
sea surface salinity observations (Level~4, https://doi.org/10.48670/moi-00051) 
and observed global mean surface temperature anomalies (version~6, https://www.ncei.noaa.gov/data/noaa-global-surface-temperature/v6/access/timeseries/)
The CMIP6 model output is provided by the World Climate Research Programme’s Working Group on Coupled Modelling.
\item Code availability -- The analysis scripts will be made available on Zenodo upon publication.
\item Authors' contributions -- R.M.v.W., E.Y.P.V. and H.A.D. conceived the idea for this study. M.K. performed the model simulation with the CESM.
R.M.v.W. conducted the analysis and prepared all figures. All authors were actively involved in the interpretation of the analysis results and the writing process.
\end{itemize}

\newpage 
\setcounter{figure}{0}

\newpage 
\setcounter{figure}{0}

\makeatletter 
\renewcommand{\thefigure}{S\@arabic\c@figure}
\renewcommand{\thetable}{S\@arabic\c@table}
\makeatother


\begin{figure}[h!]

\includegraphics[width=1\columnwidth, trim = {0cm 0cm 0cm 0cm}, clip]{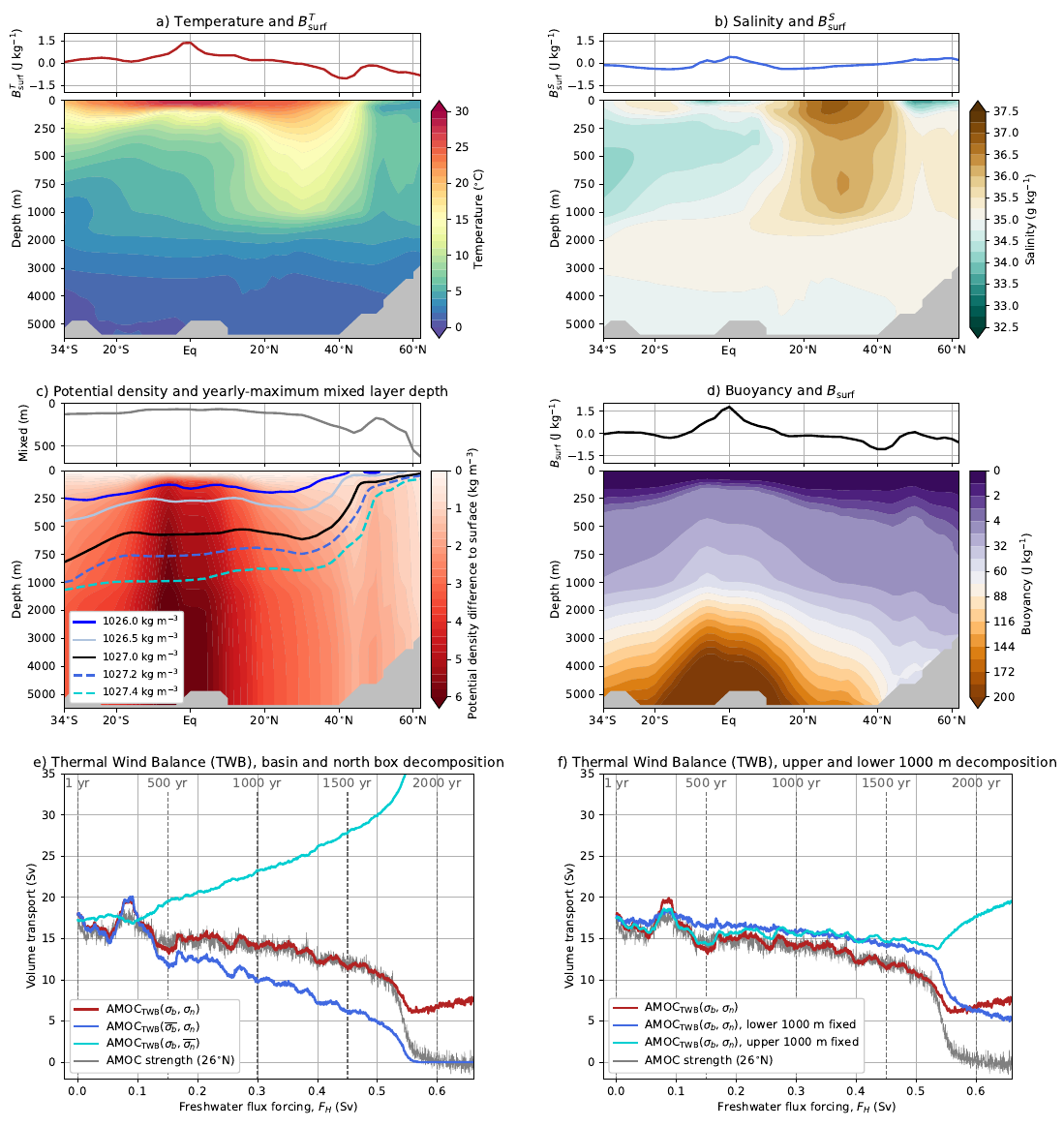}

\caption{\textbf{Atlantic Ocean water properties for the CESM (hosing).}
(a -- d): The zonally-averaged (Atlantic basin) and yearly-averaged temperature, salinity, potential density and ocean buoyancy for the first 50~years of the hosing.
The top panels show the yearly-summed surface buoyancy components and the yearly-maximum mixed layer depth.
Panel~c shows the potential density difference with respect to the surface and the curves represent various isopycnals.
(e): The thermal wind balance AMOC approximation (AMOC$_{\mathrm{TWB}}(\sigma_b,\sigma_n)$, see Methods) under the hosing.
The thermal wind balance is decomposed into a component for which the reference (i.e., first 50~model years) basin box potential density (AMOC$_{\mathrm{TWB}}(\overline{\sigma_b},\sigma_n)$) is used,
and the reference north box potential density  (AMOC$_{\mathrm{TWB}}(\sigma_b,\overline{\sigma_n})$) is used.
We also show the AMOC strength at 1,000~m and 26$^{\circ}$N for comparison.
(f): Similar to panel~e, but now the potential densities over the lower 1,000~m are fixed (i.e., first 50~model years) and the upper 1,000~m are fixed.}

\label{fig:Figure_S1}
\end{figure}


\begin{figure}[h!]

\includegraphics[width=1\columnwidth, trim = {0cm 0cm 0cm 0cm}, clip]{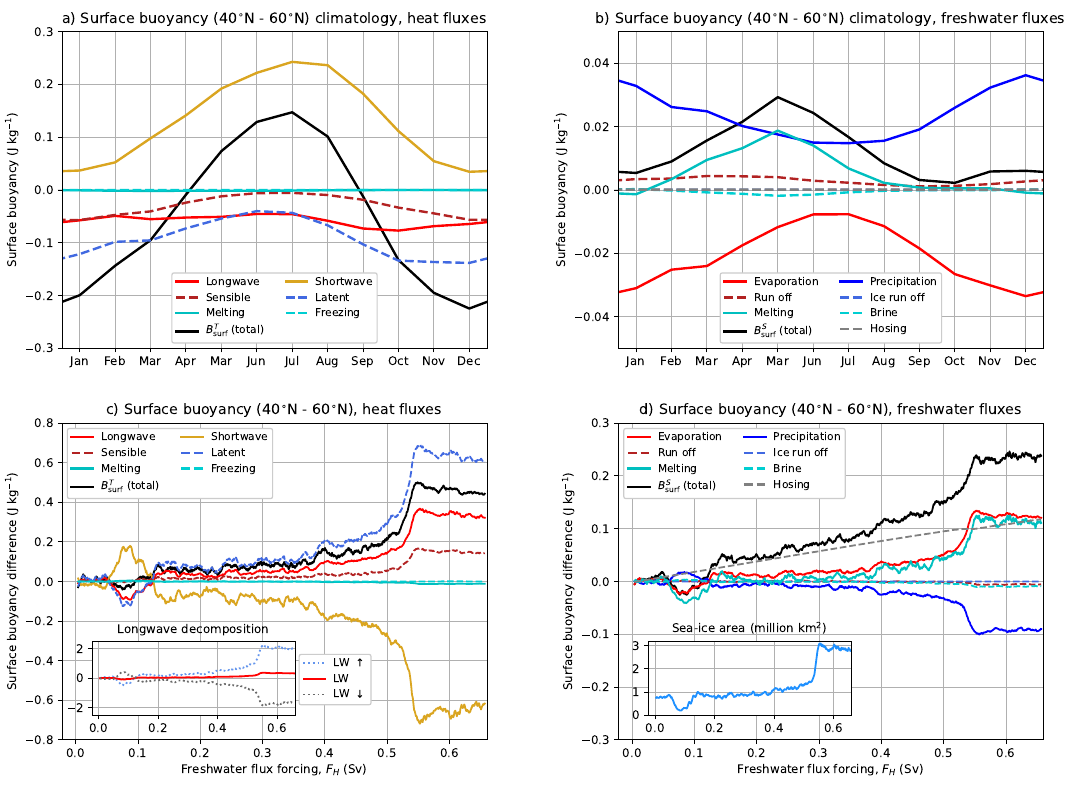}

\caption{\textbf{Surface buoyancy decomposition for the CESM (hosing).}
(a -- b): Climatology over the first 50~model years of the surface buoyancy (40$^{\circ}$N -- 60$^{\circ}$N, Atlantic Ocean) and is decomposed into its different contributions for the
(a): heat fluxes and (b): freshwater fluxes.
(c -- d): The surface buoyancy difference (with respect to the climatology) under the varying freshwater flux forcing for the (c): heat fluxes and (d): freshwater fluxes.
The time series are displayed as yearly sums and are smoothed through a 25-year running mean to reduce the variability.
The inset in panel~c shows the upwelling longwave (LW $\uparrow$) and the downwelling longwave (LW $\downarrow$) contributions. 
The inset in panel~d shows the yearly-averaged sea-ice area (grid cells with sea-ice fractions of at least 15\%) between 40$^{\circ}$N -- 60$^{\circ}$N.
Note the different vertical ranges for the panels.}

\label{fig:Figure_S2}
\end{figure}


\begin{figure}[h!]

\includegraphics[width=1\columnwidth, trim = {0cm 0cm 0cm 0cm}, clip]{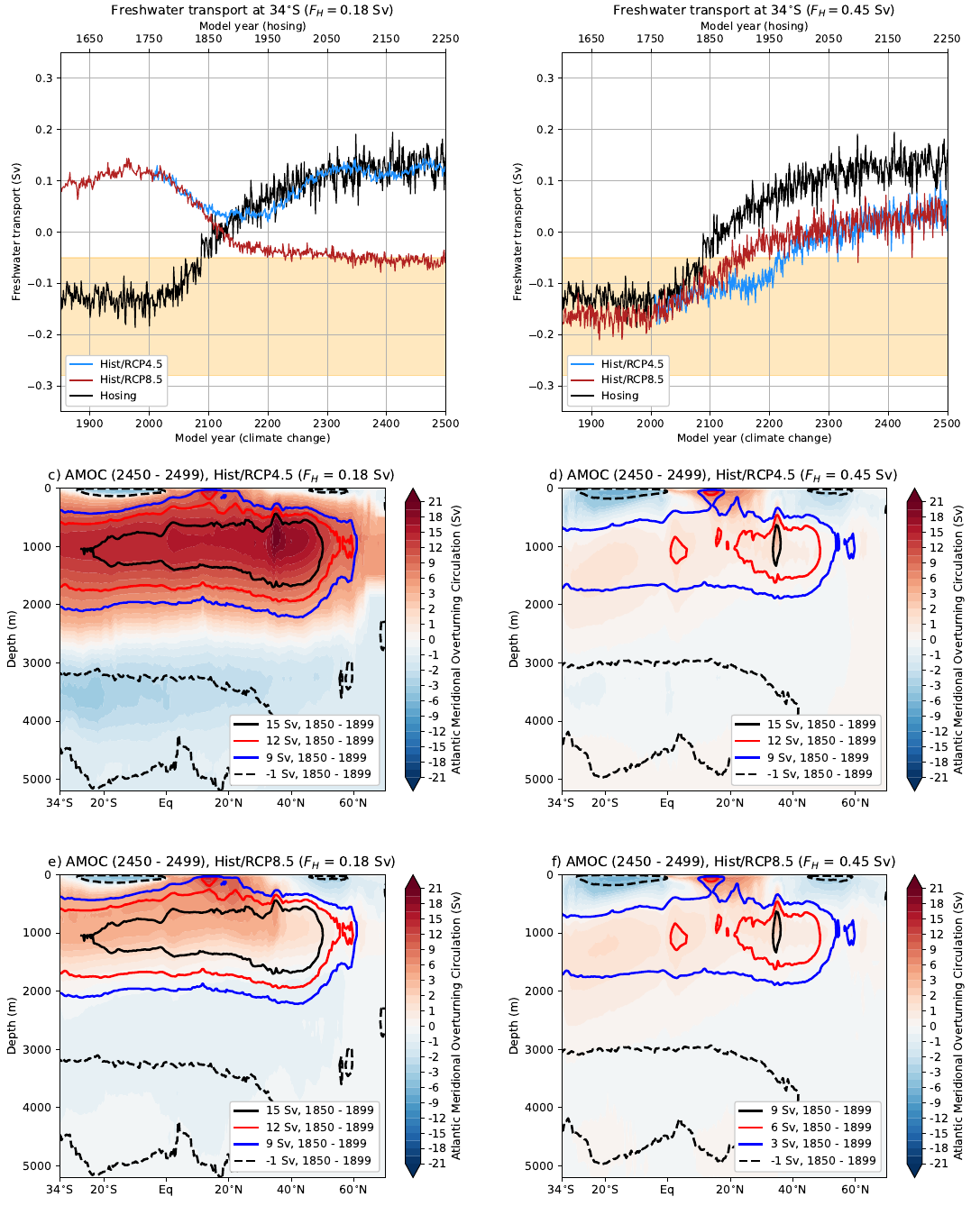}

\caption{\textbf{Atlantic Meridional Overturning Circulation (AMOC) properties.}
(a \& b): The freshwater transport carried by the AMOC at 34$^{\circ}$S, $F_{\mathrm{ovS}}$.
For reference, the $F_{\mathrm{ovS}}$ under the hosing is also shown and is centred around the AMOC tipping event (black curve, top x-axis).
The yellow shading indicates observed $F_{\mathrm{ovS}}$ values \cite{Garzoli2013, Mecking2017, ArumiPlanas2024}.
(b -- f): The AMOC streamfunction over the last 50~years (model years~2450 -- 2499).
The contours indicate the isolines  for different AMOC values for the historical period (model years~1850 -- 1899).}

\label{fig:Figure_S3}
\end{figure}


\begin{figure}[h!]

\includegraphics[width=1\columnwidth, trim = {0cm 0cm 0cm 0cm}, clip]{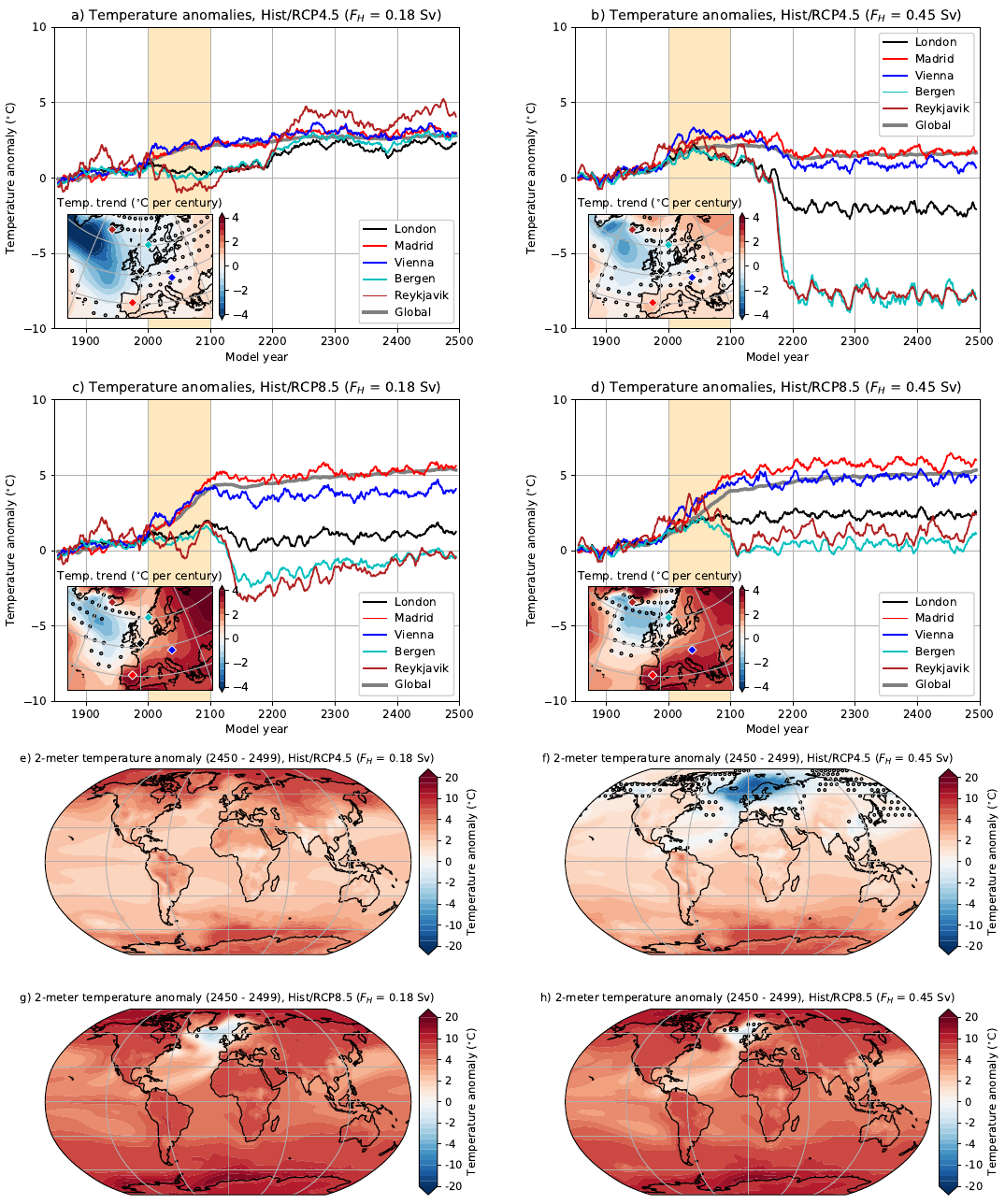}

\caption{\textbf{The surface temperature responses under climate change.}
(a -- d): The yearly-averaged surface temperatures for five different cities (diamond markers in inset) and the global mean under the two climate change scenarios 
(Hist/RCP4.5 and Hist/RCP8.5) and for $F_H = 0.18$~Sv and $F_H = 0.45$~Sv.
The time series are displayed as anomalies compared to the historical period (1850 -- 1899) and are then smoothed through a 11-year running mean. 
The yellow shading indicates the 21$^{\mathrm{st}}$ century over which the trends are displayed in the insets, the circled markers indicate non-significant ($p \geq 0.05$, \cite{Santer2000}) trends.
(e -- h): The equilibrium temperature anomalies (compared to 1850 -- 1899) at the end of the simulations,
the circled markers indicate non-significant ($p \geq 0.05$, two-sided Welch's t-test) differences.}

\label{fig:Figure_S4}
\end{figure}


\begin{figure}[h!]

\hspace{-4.5cm}
\includegraphics[width=1.6\columnwidth, trim = {0cm 0cm 0cm 0cm}, clip]{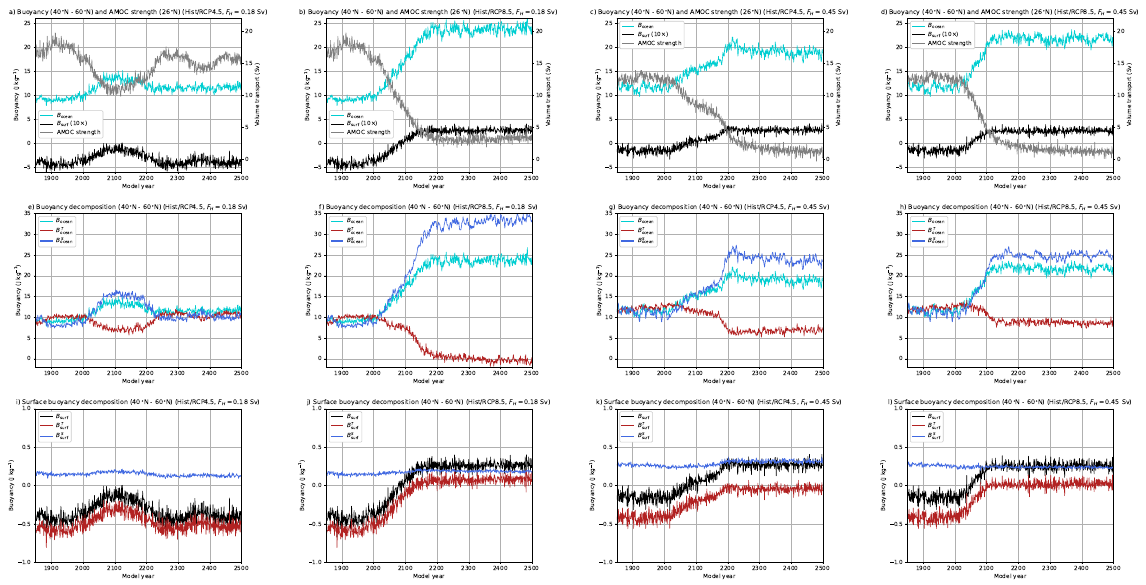}

\caption{\textbf{Buoyancy and AMOC responses under climate change.}
Similar to Figure~\ref{fig:Figure_1}, but now under the two climate change scenarios (Hist/RCP4.5 and Hist/RCP8.5) and for $F_H = 0.18$~Sv and $F_H = 0.45$~Sv.
The climatologies are shown in Figure~\ref{fig:Figure_S6}.}

\label{fig:Figure_S5}
\end{figure}


\begin{figure}[h!]

\hspace{-3.2cm}
\includegraphics[width=1.5\columnwidth, trim = {0cm 0cm 0cm 0cm}, clip]{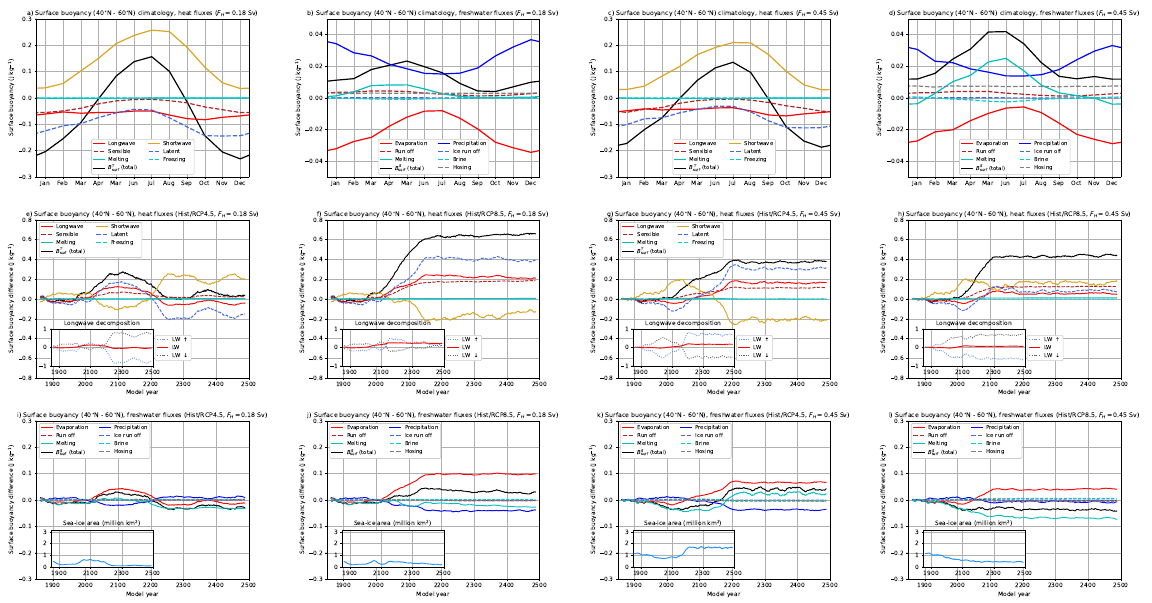}

\caption{\textbf{Surface buoyancy decomposition under climate change.}
Similar to Figure~\ref{fig:Figure_S2}, but now for the different climate change simulations.
The climatology (panels~a -- d) is determined over model years 1850 -- 1899.
The time series for the heat fluxes are shown in panels~e -- h and for the freshwater fluxes in panels~i -- l. 
Note that there are two historical simulations (for $F_H = 0.18$~Sv and $F_H = 0.45$~Sv) and hence 
the climatology and the model output between model years~1850 -- 2005 are identical when comparing the two climate change scenarios (RCP4.5 and RCP8.5).
}

\label{fig:Figure_S6}
\end{figure}


\begin{figure}[h]
\hspace{-3cm}
\includegraphics[width=1.4\columnwidth, trim = {0cm 0cm 0cm 0cm}, clip]{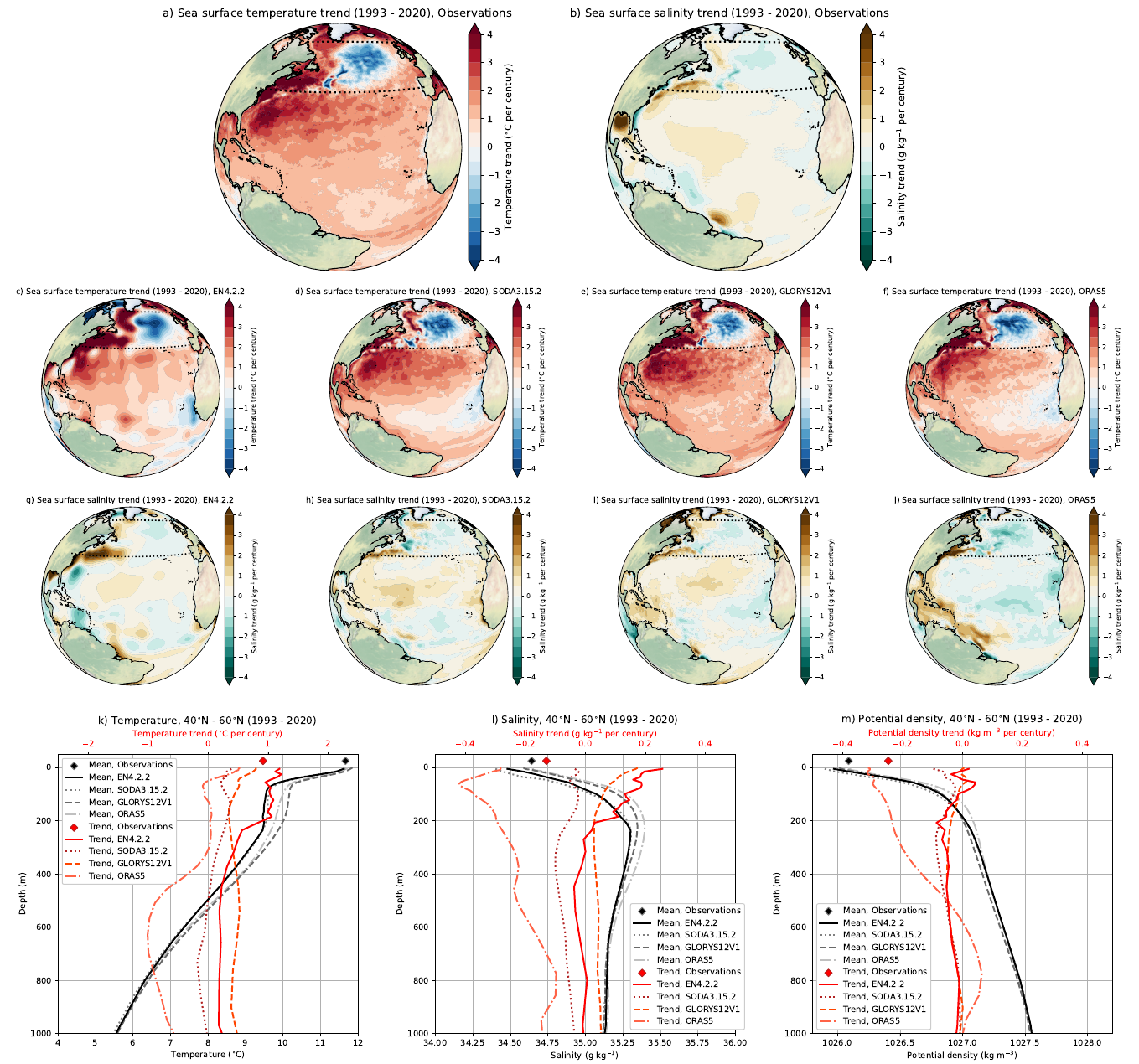}
\caption{\textbf{Present-day oceanic climate change (1993 -- 2020).}
(a -- j): The sea surface temperature and sea surface salinity trends (1993 -- 2020) for satellite observations,  EN4.2.2, SODA3.15.2, GLORYS12V1 and ORAS5.
(k -- m): The vertical structure of temperature, salinity and potential density between 40$^{\circ}$N -- 60$^{\circ}$N (Atlantic Ocean, dotted lines in panels~a -- j) and for the period 1993 -- 2020.
The black (red) curves are the time means (trends) for 1993 -- 2020 and are shown for the different products.
The diamond markers indicate the satellite observations.}
\label{fig:Figure_S7}
\end{figure}


\begin{figure}[h!]

\includegraphics[width=1\columnwidth, trim = {0cm 0cm 0cm 0cm}, clip]{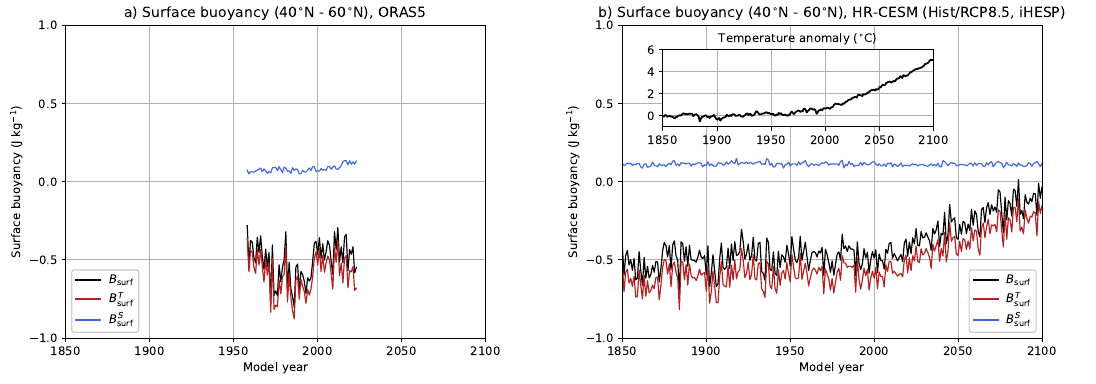}

\caption{\textbf{Surface buoyancy changes under climate change.}
The surface buoyancy ($B_{\mathrm{surf}}$) decomposition for (a): ORAS5 and 
(b): the high-resolution version of the CESM (iHESP) under the Hist/RCP8.5 scenario.
The inset in panel~b shows yearly-averaged global mean surface temperature anomaly compared to the historical period (1850 -- 1899).}

\label{fig:Figure_S8}
\end{figure}


\begin{table}
\centering

\begin{tabular}{c}

\includegraphics[width=1\columnwidth, trim = {0cm 0cm 0cm 0cm}, clip]{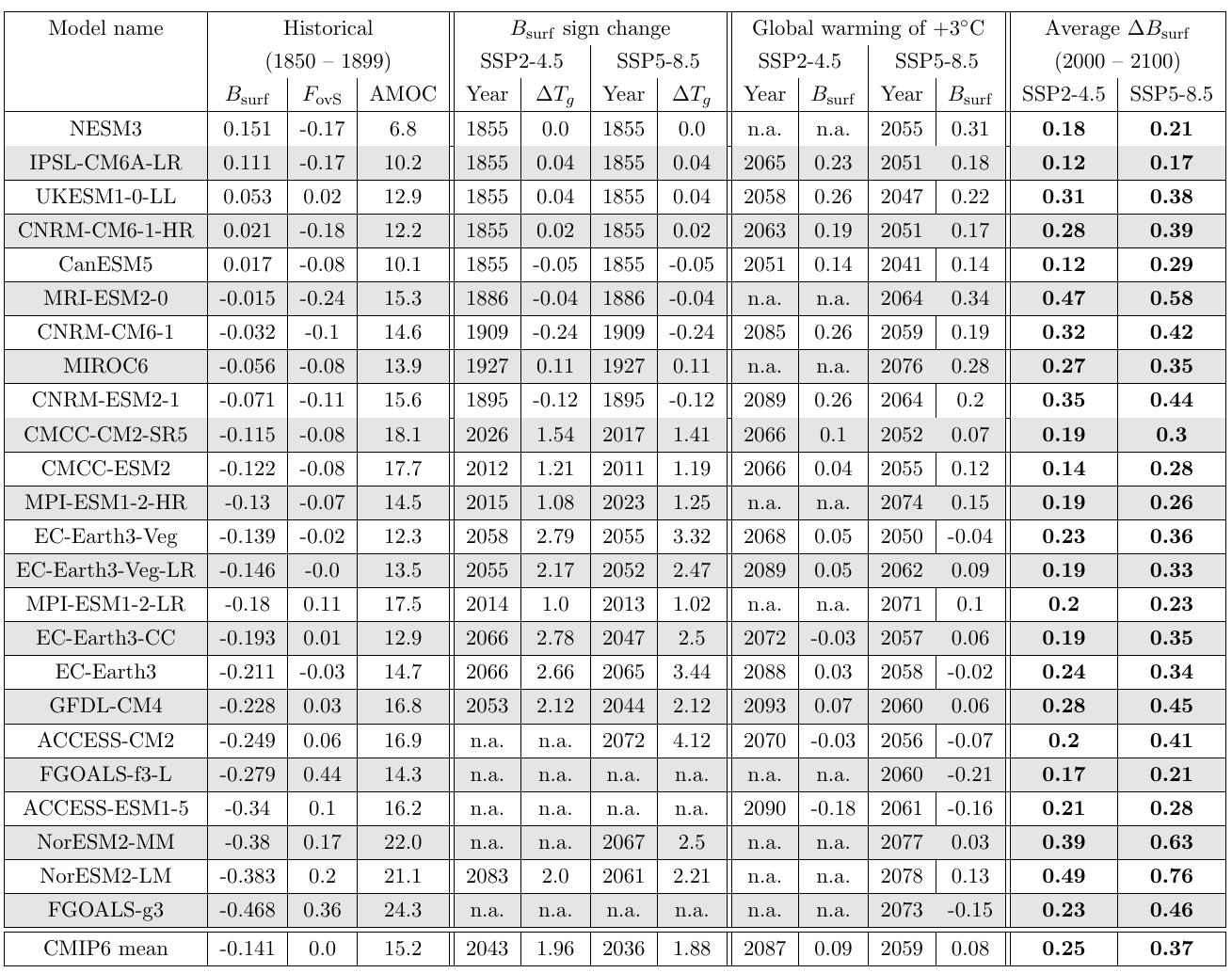}

\end{tabular}

\caption{Overview of the CMIP6 models, ranked on their historical (1850 -- 1899) surface buoyancy ($B_{\mathrm{surf}}$, in J~kg$^{-1}$) value,
including their historical freshwater transport at 34$^{\circ}$S ($F_{\mathrm{ovS}}$, in Sv) and AMOC strength at 1,000~m and 26$^{\circ}$N (in Sv).
For each CMIP6 model, we determined the first 11-year window (e.g., 2055 $\rightarrow$ 2050 -- 2060) where $B_{\mathrm{surf}}$ switches sign and the associated
global mean surface temperature anomaly ($\Delta T_g$, in $^{\circ}$C) compared to the historical period. 
Similarly, the first 11-year window where the global surface temperature anomaly exceeds 3$^{\circ}$C warming and the associated $B_{\mathrm{surf}}$ value.
The last two columns display the average $B_{\mathrm{surf}}$ change (i.e., linear trend) over the 21$^{\mathrm{st}}$ century, where bold indicates significant ($p < 0.05$, \cite{Santer2000}) changes.
For the CMIP6 mean, we first determined the multi-model mean and then determined the presented quantities.
}  

\label{tab:Table_S1}
\end{table}




\begin{thebibliography}{57}
\ifx \bisbn   \undefined \def \bisbn  #1{ISBN #1}\fi
\ifx \binits  \undefined \def \binits#1{#1}\fi
\ifx \bauthor  \undefined \def \bauthor#1{#1}\fi
\ifx \batitle  \undefined \def \batitle#1{#1}\fi
\ifx \bjtitle  \undefined \def \bjtitle#1{#1}\fi
\ifx \bvolume  \undefined \def \bvolume#1{\textbf{#1}}\fi
\ifx \byear  \undefined \def \byear#1{#1}\fi
\ifx \bissue  \undefined \def \bissue#1{#1}\fi
\ifx \bfpage  \undefined \def \bfpage#1{#1}\fi
\ifx \blpage  \undefined \def \blpage #1{#1}\fi
\ifx \burl  \undefined \def \burl#1{\textsf{#1}}\fi
\ifx \doiurl  \undefined \def \doiurl#1{\url{https://doi.org/#1}}\fi
\ifx \betal  \undefined \def \betal{\textit{et al.}}\fi
\ifx \binstitute  \undefined \def \binstitute#1{#1}\fi
\ifx \binstitutionaled  \undefined \def \binstitutionaled#1{#1}\fi
\ifx \bctitle  \undefined \def \bctitle#1{#1}\fi
\ifx \beditor  \undefined \def \beditor#1{#1}\fi
\ifx \bpublisher  \undefined \def \bpublisher#1{#1}\fi
\ifx \bbtitle  \undefined \def \bbtitle#1{#1}\fi
\ifx \bedition  \undefined \def \bedition#1{#1}\fi
\ifx \bseriesno  \undefined \def \bseriesno#1{#1}\fi
\ifx \blocation  \undefined \def \blocation#1{#1}\fi
\ifx \bsertitle  \undefined \def \bsertitle#1{#1}\fi
\ifx \bsnm \undefined \def \bsnm#1{#1}\fi
\ifx \bsuffix \undefined \def \bsuffix#1{#1}\fi
\ifx \bparticle \undefined \def \bparticle#1{#1}\fi
\ifx \barticle \undefined \def \barticle#1{#1}\fi
\bibcommenthead
\ifx \bconfdate \undefined \def \bconfdate #1{#1}\fi
\ifx \botherref \undefined \def \botherref #1{#1}\fi
\ifx \url \undefined \def \url#1{\textsf{#1}}\fi
\ifx \bchapter \undefined \def \bchapter#1{#1}\fi
\ifx \bbook \undefined \def \bbook#1{#1}\fi
\ifx \bcomment \undefined \def \bcomment#1{#1}\fi
\ifx \oauthor \undefined \def \oauthor#1{#1}\fi
\ifx \citeauthoryear \undefined \def \citeauthoryear#1{#1}\fi
\ifx \endbibitem  \undefined \def \endbibitem {}\fi
\ifx \bconflocation  \undefined \def \bconflocation#1{#1}\fi
\ifx \arxivurl  \undefined \def \arxivurl#1{\textsf{#1}}\fi
\csname PreBibitemsHook\endcsname

\bibitem[\protect\citeauthoryear{Lenton et~al.}{2008}]{Lenton2008}
\begin{barticle}
\bauthor{\bsnm{Lenton}, \binits{T.M.}},
\bauthor{\bsnm{Held}, \binits{H.}},
\bauthor{\bsnm{Kriegler}, \binits{E.}},
\bauthor{\bsnm{Hall}, \binits{J.W.}},
\bauthor{\bsnm{Lucht}, \binits{W.}},
\bauthor{\bsnm{Rahmstorf}, \binits{S.}},
\bauthor{\bsnm{Schellnhuber}, \binits{H.J.}}:
\batitle{{Tipping elements in the Earth's climate system.}}
\bjtitle{Proceedings of the National Academy of Sciences of the United States
  of America}
\bvolume{105}(\bissue{6}),
\bfpage{1786}--\blpage{93}
(\byear{2008})
\doiurl{10.1073/pnas.0705414105}
\end{barticle}
\endbibitem

\bibitem[\protect\citeauthoryear{Armstrong~McKay et~al.}{2022}]{Armstrong2022}
\begin{barticle}
\bauthor{\bsnm{Armstrong~McKay}, \binits{D.I.}},
\bauthor{\bsnm{Staal}, \binits{A.}},
\bauthor{\bsnm{Abrams}, \binits{J.F.}},
\bauthor{\bsnm{Winkelmann}, \binits{R.}},
\bauthor{\bsnm{Sakschewski}, \binits{B.}},
\bauthor{\bsnm{Loriani}, \binits{S.}},
\bauthor{\bsnm{Fetzer}, \binits{I.}},
\bauthor{\bsnm{Cornell}, \binits{S.E.}},
\bauthor{\bsnm{Rockstr{\"o}m}, \binits{J.}},
\bauthor{\bsnm{Lenton}, \binits{T.M.}}:
\batitle{{Exceeding 1.5 C global warming could trigger multiple climate tipping
  points}}.
\bjtitle{Science}
\bvolume{377}(\bissue{6611}),
\bfpage{7950}
(\byear{2022})
\end{barticle}
\endbibitem

\bibitem[\protect\citeauthoryear{Johns et~al.}{2011}]{Johns2011}
\begin{barticle}
\bauthor{\bsnm{Johns}, \binits{W.E.}},
\bauthor{\bsnm{Baringer}, \binits{M.O.}},
\bauthor{\bsnm{Beal}, \binits{L.}},
\bauthor{\bsnm{Cunningham}, \binits{S.}},
\bauthor{\bsnm{Kanzow}, \binits{T.}},
\bauthor{\bsnm{Bryden}, \binits{H.L.}},
\bauthor{\bsnm{Hirschi}, \binits{J.}},
\bauthor{\bsnm{Marotzke}, \binits{J.}},
\bauthor{\bsnm{Meinen}, \binits{C.}},
\bauthor{\bsnm{Shaw}, \binits{B.}}, \betal:
\batitle{{Continuous, array-based estimates of Atlantic Ocean heat transport at
  26.5 N}}.
\bjtitle{Journal of Climate}
\bvolume{24}(\bissue{10}),
\bfpage{2429}--\blpage{2449}
(\byear{2011})
\end{barticle}
\endbibitem

\bibitem[\protect\citeauthoryear{Orihuela-Pinto et~al.}{2022}]{Orihuela2022}
\begin{barticle}
\bauthor{\bsnm{Orihuela-Pinto}, \binits{B.}},
\bauthor{\bsnm{England}, \binits{M.H.}},
\bauthor{\bsnm{Taschetto}, \binits{A.S.}}:
\batitle{Interbasin and interhemispheric impacts of a collapsed atlantic
  overturning circulation}.
\bjtitle{Nature Climate Change}
\bvolume{12}(\bissue{6}),
\bfpage{558}--\blpage{565}
(\byear{2022})
\end{barticle}
\endbibitem

\bibitem[\protect\citeauthoryear{van Westen and
  Dijkstra}{2023}]{vanWesten2023b}
\begin{barticle}
\bauthor{\bsnm{Westen}, \binits{R.M.}},
\bauthor{\bsnm{Dijkstra}, \binits{H.A.}}:
\batitle{{Asymmetry of AMOC Hysteresis in a State-Of-The-Art Global Climate
  Model}}.
\bjtitle{Geophysical Research Letters}
\bvolume{50}(\bissue{22}),
\bfpage{2023}--\blpage{106088}
(\byear{2023})
\end{barticle}
\endbibitem

\bibitem[\protect\citeauthoryear{van Westen et~al.}{2024}]{vanWesten2024a}
\begin{barticle}
\bauthor{\bsnm{Westen}, \binits{R.M.}},
\bauthor{\bsnm{Kliphuis}, \binits{M.}},
\bauthor{\bsnm{Dijkstra}, \binits{H.A.}}:
\batitle{{Physics-based early warning signal shows that AMOC is on tipping
  course}}.
\bjtitle{Science advances}
\bvolume{10}(\bissue{6}),
\bfpage{1189}
(\byear{2024})
\end{barticle}
\endbibitem

\bibitem[\protect\citeauthoryear{Garzoli et~al.}{2013}]{Garzoli2013}
\begin{barticle}
\bauthor{\bsnm{Garzoli}, \binits{S.L.}},
\bauthor{\bsnm{Baringer}, \binits{M.O.}},
\bauthor{\bsnm{Dong}, \binits{S.}},
\bauthor{\bsnm{Perez}, \binits{R.C.}},
\bauthor{\bsnm{Yao}, \binits{Q.}}:
\batitle{{South Atlantic meridional fluxes}}.
\bjtitle{Deep Sea Research Part I: Oceanographic Research Papers}
\bvolume{71},
\bfpage{21}--\blpage{32}
(\byear{2013})
\end{barticle}
\endbibitem

\bibitem[\protect\citeauthoryear{Srokosz and Bryden}{2015}]{Srokosz2015}
\begin{barticle}
\bauthor{\bsnm{Srokosz}, \binits{M.A.}},
\bauthor{\bsnm{Bryden}, \binits{H.L.}}:
\batitle{{Observing the Atlantic Meridional Overturning Circulation yields a
  decade of inevitable surprises.}}
\bjtitle{Science}
\bvolume{348}(\bissue{6241}),
\bfpage{1255575}--\blpage{1255575}
(\byear{2015})
\doiurl{10.1126/science.1255575}
\end{barticle}
\endbibitem

\bibitem[\protect\citeauthoryear{Lozier et~al.}{2019}]{Lozier2019}
\begin{barticle}
\bauthor{\bsnm{Lozier}, \binits{M.S.}},
\bauthor{\bsnm{Li}, \binits{F.}},
\bauthor{\bsnm{Bacon}, \binits{S.}},
\bauthor{\bsnm{Bahr}, \binits{F.}},
\bauthor{\bsnm{Bower}, \binits{A.S.}},
\bauthor{\bsnm{Cunningham}, \binits{S.}},
\bauthor{\bsnm{Jong}, \binits{M.F.}},
\bauthor{\bsnm{Steur}, \binits{L.}},
\bauthor{\bsnm{deYoung}, \binits{B.}},
\bauthor{\bsnm{Fischer}, \binits{J.}}, \betal:
\batitle{A sea change in our view of overturning in the subpolar north
  atlantic}.
\bjtitle{Science}
\bvolume{363}(\bissue{6426}),
\bfpage{516}--\blpage{521}
(\byear{2019})
\end{barticle}
\endbibitem

\bibitem[\protect\citeauthoryear{Caesar et~al.}{2018}]{Ceasar2018}
\begin{barticle}
\bauthor{\bsnm{Caesar}, \binits{L.}},
\bauthor{\bsnm{Rahmstorf}, \binits{S.}},
\bauthor{\bsnm{Robinson}, \binits{A.}},
\bauthor{\bsnm{Feulner}, \binits{G.}},
\bauthor{\bsnm{Saba}, \binits{V.}}:
\batitle{{Observed fingerprint of a weakening Atlantic Ocean overturning
  circulation}}.
\bjtitle{Nature}
\bvolume{556}(\bissue{7700}),
\bfpage{191}--\blpage{196}
(\byear{2018})
\doiurl{10.1038/s41586-018-0006-5}
\end{barticle}
\endbibitem

\bibitem[\protect\citeauthoryear{Estella-Perez
  et~al.}{2020}]{Estella-Perez2020}
\begin{barticle}
\bauthor{\bsnm{Estella-Perez}, \binits{V.}},
\bauthor{\bsnm{Mignot}, \binits{J.}},
\bauthor{\bsnm{Guilyardi}, \binits{E.}},
\bauthor{\bsnm{Swingedouw}, \binits{D.}},
\bauthor{\bsnm{Reverdin}, \binits{G.}}:
\batitle{Advances in reconstructing the amoc using sea surface observations of
  salinity}.
\bjtitle{Climate Dynamics}
\bvolume{55},
\bfpage{975}--\blpage{992}
(\byear{2020})
\end{barticle}
\endbibitem

\bibitem[\protect\citeauthoryear{Caesar et~al.}{2021}]{Ceasar2021}
\begin{barticle}
\bauthor{\bsnm{Caesar}, \binits{L.}},
\bauthor{\bsnm{McCarthy}, \binits{G.D.}},
\bauthor{\bsnm{Thornalley}, \binits{D.J.R.}},
\bauthor{\bsnm{Cahill}, \binits{N.}},
\bauthor{\bsnm{Rahmstorf}, \binits{S.}}:
\batitle{{Current Atlantic Meridional Overturning Circulation weakest in last
  millennium}}.
\bjtitle{Nature Geoscience}
\bvolume{14}(\bissue{3}),
\bfpage{118}--\blpage{120}
(\byear{2021})
\doiurl{10.1038/s41561-021-00699-z}
\end{barticle}
\endbibitem

\bibitem[\protect\citeauthoryear{Worthington et~al.}{2021}]{Worthington2021}
\begin{barticle}
\bauthor{\bsnm{Worthington}, \binits{E.L.}},
\bauthor{\bsnm{Moat}, \binits{B.I.}},
\bauthor{\bsnm{Smeed}, \binits{D.A.}},
\bauthor{\bsnm{Mecking}, \binits{J.V.}},
\bauthor{\bsnm{Marsh}, \binits{R.}},
\bauthor{\bsnm{McCarthy}, \binits{G.D.}}:
\batitle{{A 30-year reconstruction of the Atlantic meridional overturning
  circulation shows no decline}}.
\bjtitle{Ocean Science}
\bvolume{17}(\bissue{1}),
\bfpage{285}--\blpage{299}
(\byear{2021})
\end{barticle}
\endbibitem

\bibitem[\protect\citeauthoryear{Michel et~al.}{2023}]{Michel2023}
\begin{botherref}
\oauthor{\bsnm{Michel}, \binits{S.}},
\oauthor{\bsnm{Dijkstra}, \binits{H.}},
\oauthor{\bsnm{Guardamagna}, \binits{F.}},
\oauthor{\bsnm{Jacques-Dumas}, \binits{V.}},
\oauthor{\bsnm{Westen}, \binits{R.}},
\oauthor{\bsnm{Heydt}, \binits{A.}}:
Deep learning reconstruction of atlantic meridional overturning circulation
  strength validates ongoig twenty-first century decline.
(2023)
\end{botherref}
\endbibitem

\bibitem[\protect\citeauthoryear{Boers}{2021}]{Boers2021}
\begin{barticle}
\bauthor{\bsnm{Boers}, \binits{N.}}:
\batitle{{Observation-based early-warning signals for a collapse of the
  Atlantic Meridional Overturning Circulation}}.
\bjtitle{Nature Climate Change}
\bvolume{11}(\bissue{8}),
\bfpage{680}--\blpage{688}
(\byear{2021})
\doiurl{10.1038/s41558-021-01097-4}
\end{barticle}
\endbibitem

\bibitem[\protect\citeauthoryear{Ditlevsen and Ditlevsen}{2023}]{Ditlevsen2023}
\begin{barticle}
\bauthor{\bsnm{Ditlevsen}, \binits{P.}},
\bauthor{\bsnm{Ditlevsen}, \binits{S.}}:
\batitle{{Warning of a forthcoming collapse of the Atlantic meridional
  overturning circulation}}.
\bjtitle{Nature Communications}
\bvolume{14}(\bissue{1}),
\bfpage{4254}
(\byear{2023})
\end{barticle}
\endbibitem

\bibitem[\protect\citeauthoryear{Smolders et~al.}{2024}]{Smolders2024}
\begin{botherref}
\oauthor{\bsnm{Smolders}, \binits{E.J.V.}},
\oauthor{\bsnm{Westen}, \binits{R.M.}},
\oauthor{\bsnm{Dijkstra}, \binits{H.A.}}:
Probability estimates of a 21st century amoc collapse
(2024)
{\href{https://arxiv.org/abs/2406.11738}{{arXiv:2406.11738}}}
{[physics.ao-ph]}
\end{botherref}
\endbibitem

\bibitem[\protect\citeauthoryear{Weijer et~al.}{2020}]{Weijer2020}
\begin{barticle}
\bauthor{\bsnm{Weijer}, \binits{W.}},
\bauthor{\bsnm{Cheng}, \binits{W.}},
\bauthor{\bsnm{Garuba}, \binits{O.A.}},
\bauthor{\bsnm{Hu}, \binits{A.}},
\bauthor{\bsnm{Nadiga}, \binits{B.T.}}:
\batitle{{CMIP6 Models Predict Significant 21st Century Decline of the Atlantic
  Meridional Overturning Circulation}}.
\bjtitle{Geophysical Research Letters}
\bvolume{47}(\bissue{12}),
\bfpage{2019}--\blpage{08607}
(\byear{2020})
\doiurl{10.1029/2019gl086075}
\end{barticle}
\endbibitem

\bibitem[\protect\citeauthoryear{Fox-Kemper et~al.}{2021}]{Fox-Kemper2021}
\begin{botherref}
\oauthor{\bsnm{Fox-Kemper}, \binits{B.}},
\oauthor{\bsnm{Hewitt}, \binits{H.T.}},
\oauthor{\bsnm{Xiao}, \binits{C.}},
\oauthor{\bsnm{co-authors}}:
Ocean, cryosphere and sea level change. in climate change 2021: The physical
  science basis. contribution of working group i to the sixth assessment report
  of the intergovernmental panel on climate change,
1211--1362
(2021)
\doiurl{10.1017/9781009157896.011}
\end{botherref}
\endbibitem

\bibitem[\protect\citeauthoryear{Nikurashin and Vallis}{2012}]{Nikurashin2012}
\begin{barticle}
\bauthor{\bsnm{Nikurashin}, \binits{M.}},
\bauthor{\bsnm{Vallis}, \binits{G.}}:
\batitle{A theory of the interhemispheric meridional overturning circulation
  and associated stratification}.
\bjtitle{Journal of Physical Oceanography}
\bvolume{42}(\bissue{10}),
\bfpage{1652}--\blpage{1667}
(\byear{2012})
\end{barticle}
\endbibitem

\bibitem[\protect\citeauthoryear{Wolfe and Cessi}{2014}]{Wolfe2014}
\begin{barticle}
\bauthor{\bsnm{Wolfe}, \binits{C.L.}},
\bauthor{\bsnm{Cessi}, \binits{P.}}:
\batitle{Salt feedback in the adiabatic overturning circulation}.
\bjtitle{Journal of Physical Oceanography}
\bvolume{44}(\bissue{4}),
\bfpage{1175}--\blpage{1194}
(\byear{2014})
\end{barticle}
\endbibitem

\bibitem[\protect\citeauthoryear{Wolfe and Cessi}{2015}]{Wolfe2015}
\begin{barticle}
\bauthor{\bsnm{Wolfe}, \binits{C.L.}},
\bauthor{\bsnm{Cessi}, \binits{P.}}:
\batitle{Multiple regimes and low-frequency variability in the quasi-adiabatic
  overturning circulation}.
\bjtitle{Journal of Physical Oceanography}
\bvolume{45}(\bissue{6}),
\bfpage{1690}--\blpage{1708}
(\byear{2015})
\end{barticle}
\endbibitem

\bibitem[\protect\citeauthoryear{Walin}{1982}]{Walin1982}
\begin{barticle}
\bauthor{\bsnm{Walin}, \binits{G.}}:
\batitle{On the relation between sea-surface heat flow and thermal circulation
  in the ocean}.
\bjtitle{Tellus}
\bvolume{34}(\bissue{2}),
\bfpage{187}--\blpage{195}
(\byear{1982})
\end{barticle}
\endbibitem

\bibitem[\protect\citeauthoryear{Marshall et~al.}{1999}]{Marshall1999}
\begin{barticle}
\bauthor{\bsnm{Marshall}, \binits{J.}},
\bauthor{\bsnm{Jamous}, \binits{D.}},
\bauthor{\bsnm{Nilsson}, \binits{J.}}:
\batitle{{Reconciling thermodynamic and dynamic methods of computation of
  water-mass transformation rates}}.
\bjtitle{Deep Sea Research Part I: Oceanographic Research Papers}
\bvolume{46}(\bissue{4}),
\bfpage{545}--\blpage{572}
(\byear{1999})
\end{barticle}
\endbibitem

\bibitem[\protect\citeauthoryear{Marotzke}{2000}]{Marotzke2000}
\begin{barticle}
\bauthor{\bsnm{Marotzke}, \binits{J.}}:
\batitle{{Abrupt climate change and thermohaline circulation: Mechanisms and
  Predictability}}.
\bjtitle{Proc. Natl. Acad. Sci.}
\bvolume{97},
\bfpage{1347}--\blpage{1350}
(\byear{2000})
\end{barticle}
\endbibitem

\bibitem[\protect\citeauthoryear{Peltier and Vettoretti}{2014}]{Peltier2014}
\begin{barticle}
\bauthor{\bsnm{Peltier}, \binits{W.R.}},
\bauthor{\bsnm{Vettoretti}, \binits{G.}}:
\batitle{{Dansgaard-Oeschger oscillations predicted in a comprehensive model of
  glacial climate: A “kicked” salt oscillator in the Atlantic}}.
\bjtitle{Geophysical Research Letters}
\bvolume{41}(\bissue{20}),
\bfpage{7306}--\blpage{7313}
(\byear{2014})
\end{barticle}
\endbibitem

\bibitem[\protect\citeauthoryear{Lin et~al.}{2023}]{Lin2023}
\begin{barticle}
\bauthor{\bsnm{Lin}, \binits{Y.-J.}},
\bauthor{\bsnm{Rose}, \binits{B.E.}},
\bauthor{\bsnm{Hwang}, \binits{Y.-T.}}:
\batitle{{Mean state AMOC affects AMOC weakening through subsurface warming in
  the Labrador Sea}}.
\bjtitle{Journal of Climate}
\bvolume{36}(\bissue{12}),
\bfpage{3895}--\blpage{3915}
(\byear{2023})
\end{barticle}
\endbibitem

\bibitem[\protect\citeauthoryear{van Westen et~al.}{2024}]{vanWesten2024c}
\begin{botherref}
\oauthor{\bsnm{Westen}, \binits{R.M.}},
\oauthor{\bsnm{Jacques-Dumas}, \binits{V.}},
\oauthor{\bsnm{Boot}, \binits{A.A.}},
\oauthor{\bsnm{Dijkstra}, \binits{H.A.}}:
{The Role of Sea-ice Processes on the Probability of AMOC Transitions}.
https://doi.org/10.48550/arXiv.2401.12615
(2024)
\end{botherref}
\endbibitem

\bibitem[\protect\citeauthoryear{Smeed et~al.}{2018}]{Smeed2018}
\begin{barticle}
\bauthor{\bsnm{Smeed}, \binits{D.A.}},
\bauthor{\bsnm{Josey}, \binits{S.}},
\bauthor{\bsnm{Beaulieu}, \binits{C.}},
\bauthor{\bsnm{Johns}, \binits{W.}},
\bauthor{\bsnm{Moat}, \binits{B.I.}},
\bauthor{\bsnm{Frajka-Williams}, \binits{E.}},
\bauthor{\bsnm{Rayner}, \binits{D.}},
\bauthor{\bsnm{Meinen}, \binits{C.S.}},
\bauthor{\bsnm{Baringer}, \binits{M.O.}},
\bauthor{\bsnm{Bryden}, \binits{H.L.}}, \betal:
\batitle{{The North Atlantic Ocean is in a state of reduced overturning}}.
\bjtitle{Geophysical Research Letters}
\bvolume{45}(\bissue{3}),
\bfpage{1527}--\blpage{1533}
(\byear{2018})
\end{barticle}
\endbibitem

\bibitem[\protect\citeauthoryear{Dijkstra}{2007}]{Dijkstra2007}
\begin{barticle}
\bauthor{\bsnm{Dijkstra}, \binits{H.A.}}:
\batitle{Characterization of the multiple equilibria regime in a global ocean
  model}.
\bjtitle{Tellus A: Dynamic Meteorology and Oceanography}
\bvolume{59}(\bissue{5}),
\bfpage{695}--\blpage{705}
(\byear{2007})
\end{barticle}
\endbibitem

\bibitem[\protect\citeauthoryear{Huisman et~al.}{2010}]{Huisman2010}
\begin{barticle}
\bauthor{\bsnm{Huisman}, \binits{S.E.}},
\bauthor{\bsnm{Den~Toom}, \binits{M.}},
\bauthor{\bsnm{Dijkstra}, \binits{H.A.}},
\bauthor{\bsnm{Drijfhout}, \binits{S.}}:
\batitle{An indicator of the multiple equilibria regime of the atlantic
  meridional overturning circulation}.
\bjtitle{Journal of Physical Oceanography}
\bvolume{40}(\bissue{3}),
\bfpage{551}--\blpage{567}
(\byear{2010})
\end{barticle}
\endbibitem

\bibitem[\protect\citeauthoryear{Mecking et~al.}{2017}]{Mecking2017}
\begin{barticle}
\bauthor{\bsnm{Mecking}, \binits{J.}},
\bauthor{\bsnm{Drijfhout}, \binits{S.}},
\bauthor{\bsnm{Jackson}, \binits{L.}},
\bauthor{\bsnm{Andrews}, \binits{M.}}:
\batitle{{The effect of model bias on Atlantic freshwater transport and
  implications for AMOC bi-stability}}.
\bjtitle{Tellus A: Dynamic Meteorology and Oceanography}
\bvolume{69}(\bissue{1}),
\bfpage{1299910}
(\byear{2017})
\end{barticle}
\endbibitem

\bibitem[\protect\citeauthoryear{Weijer et~al.}{2019}]{Weijer2019}
\begin{barticle}
\bauthor{\bsnm{Weijer}, \binits{W.}},
\bauthor{\bsnm{Cheng}, \binits{W.}},
\bauthor{\bsnm{Drijfhout}, \binits{S.S.}},
\bauthor{\bsnm{Fedorov}, \binits{A.V.}},
\bauthor{\bsnm{Hu}, \binits{A.}},
\bauthor{\bsnm{Jackson}, \binits{L.C.}},
\bauthor{\bsnm{Liu}, \binits{W.}},
\bauthor{\bsnm{McDonagh}, \binits{E.}},
\bauthor{\bsnm{Mecking}, \binits{J.}},
\bauthor{\bsnm{Zhang}, \binits{J.}}:
\batitle{{Stability of the Atlantic Meridional Overturning Circulation: A
  review and synthesis}}.
\bjtitle{Journal of Geophysical Research: Oceans}
\bvolume{124}(\bissue{8}),
\bfpage{5336}--\blpage{5375}
(\byear{2019})
\end{barticle}
\endbibitem

\bibitem[\protect\citeauthoryear{Jackson et~al.}{2022}]{Jackson2022b}
\begin{barticle}
\bauthor{\bsnm{Jackson}, \binits{L.C.}},
\bauthor{\bsnm{Asenjo}, \binits{E.}},
\bauthor{\bsnm{Bellomo}, \binits{K.}},
\bauthor{\bsnm{Danabasoglu}, \binits{G.}},
\bauthor{\bsnm{Haak}, \binits{H.}},
\bauthor{\bsnm{Hu}, \binits{A.}},
\bauthor{\bsnm{Jungclaus}, \binits{J.}},
\bauthor{\bsnm{Lee}, \binits{W.}},
\bauthor{\bsnm{Meccia}, \binits{V.L.}},
\bauthor{\bsnm{Saenko}, \binits{O.}}, \betal:
\batitle{{Understanding AMOC stability: the North Atlantic hosing model
  intercomparison project}}.
\bjtitle{Geoscientific Model Development Discussions}
\bvolume{2022},
\bfpage{1}--\blpage{32}
(\byear{2022})
\end{barticle}
\endbibitem

\bibitem[\protect\citeauthoryear{Santer et~al.}{2000}]{Santer2000}
\begin{barticle}
\bauthor{\bsnm{Santer}, \binits{B.D.}},
\bauthor{\bsnm{Wigley}, \binits{T.}},
\bauthor{\bsnm{Boyle}, \binits{J.}},
\bauthor{\bsnm{Gaffen}, \binits{D.J.}},
\bauthor{\bsnm{Hnilo}, \binits{J.}},
\bauthor{\bsnm{Nychka}, \binits{D.}},
\bauthor{\bsnm{Parker}, \binits{D.}},
\bauthor{\bsnm{Taylor}, \binits{K.}}:
\batitle{{Statistical significance of trends and trend differences in
  layer-average atmospheric temperature time series}}.
\bjtitle{Journal of Geophysical Research: Atmospheres}
\bvolume{105}(\bissue{D6}),
\bfpage{7337}--\blpage{7356}
(\byear{2000})
\end{barticle}
\endbibitem

\bibitem[\protect\citeauthoryear{Van~Westen and
  Dijkstra}{2024}]{vanWesten2024b}
\begin{barticle}
\bauthor{\bsnm{Van~Westen}, \binits{R.M.}},
\bauthor{\bsnm{Dijkstra}, \binits{H.A.}}:
\batitle{Persistent climate model biases in the atlantic ocean's freshwater
  transport}.
\bjtitle{Ocean Science}
\bvolume{20}(\bissue{2}),
\bfpage{549}--\blpage{567}
(\byear{2024})
\end{barticle}
\endbibitem

\bibitem[\protect\citeauthoryear{Liu et~al.}{2017}]{Liu2017}
\begin{barticle}
\bauthor{\bsnm{Liu}, \binits{W.}},
\bauthor{\bsnm{Xie}, \binits{S.-P.}},
\bauthor{\bsnm{Liu}, \binits{Z.}},
\bauthor{\bsnm{Zhu}, \binits{J.}}:
\batitle{{Overlooked possibility of a collapsed Atlantic Meridional Overturning
  Circulation in warming climate}}.
\bjtitle{Science Advances}
\bvolume{3}(\bissue{1}),
\bfpage{1601666}
(\byear{2017})
\doiurl{10.1126/sciadv.1601666}
\end{barticle}
\endbibitem

\bibitem[\protect\citeauthoryear{Levang and Schmitt}{2020}]{Levang2020}
\begin{barticle}
\bauthor{\bsnm{Levang}, \binits{S.J.}},
\bauthor{\bsnm{Schmitt}, \binits{R.W.}}:
\batitle{{What causes the AMOC to weaken in CMIP5?}}
\bjtitle{Journal of Climate}
\bvolume{33}(\bissue{4}),
\bfpage{1535}--\blpage{1545}
(\byear{2020})
\end{barticle}
\endbibitem

\bibitem[\protect\citeauthoryear{Bonan et~al.}{2022}]{Bonan2022}
\begin{barticle}
\bauthor{\bsnm{Bonan}, \binits{D.B.}},
\bauthor{\bsnm{Thompson}, \binits{A.F.}},
\bauthor{\bsnm{Newsom}, \binits{E.R.}},
\bauthor{\bsnm{Sun}, \binits{S.}},
\bauthor{\bsnm{Rugenstein}, \binits{M.}}:
\batitle{{Transient and equilibrium responses of the Atlantic overturning
  circulation to warming in coupled climate models: The role of temperature and
  salinity}}.
\bjtitle{Journal of Climate}
\bvolume{35}(\bissue{15}),
\bfpage{5173}--\blpage{5193}
(\byear{2022})
\end{barticle}
\endbibitem

\bibitem[\protect\citeauthoryear{Little et~al.}{2020}]{Little2020}
\begin{barticle}
\bauthor{\bsnm{Little}, \binits{C.M.}},
\bauthor{\bsnm{Zhao}, \binits{M.}},
\bauthor{\bsnm{Buckley}, \binits{M.W.}}:
\batitle{{Do surface temperature indices reflect centennial-timescale trends in
  Atlantic meridional overturning circulation strength?}}
\bjtitle{Geophysical Research Letters}
\bvolume{47}(\bissue{22}),
\bfpage{2020}--\blpage{090888}
(\byear{2020})
\end{barticle}
\endbibitem

\bibitem[\protect\citeauthoryear{He et~al.}{2022}]{He2022}
\begin{barticle}
\bauthor{\bsnm{He}, \binits{C.}},
\bauthor{\bsnm{Clement}, \binits{A.C.}},
\bauthor{\bsnm{Cane}, \binits{M.A.}},
\bauthor{\bsnm{Murphy}, \binits{L.N.}},
\bauthor{\bsnm{Klavans}, \binits{J.M.}},
\bauthor{\bsnm{Fenske}, \binits{T.M.}}:
\batitle{{A North Atlantic warming hole without ocean circulation}}.
\bjtitle{Geophysical research letters}
\bvolume{49}(\bissue{19}),
\bfpage{2022}--\blpage{100420}
(\byear{2022})
\end{barticle}
\endbibitem

\bibitem[\protect\citeauthoryear{G{\'e}rard and Crucifix}{2024}]{Gerard2024}
\begin{barticle}
\bauthor{\bsnm{G{\'e}rard}, \binits{J.}},
\bauthor{\bsnm{Crucifix}, \binits{M.}}:
\batitle{{Diagnosing the causes of AMOC slowdown in a coupled model: a
  cautionary tale}}.
\bjtitle{Earth System Dynamics}
\bvolume{15}(\bissue{2}),
\bfpage{293}--\blpage{306}
(\byear{2024})
\end{barticle}
\endbibitem

\bibitem[\protect\citeauthoryear{Jackson}{2013}]{Jackson2013}
\begin{barticle}
\bauthor{\bsnm{Jackson}, \binits{L.C.}}:
\batitle{{Shutdown and recovery of the AMOC in a coupled global climate model:
  The role of the advective feedback}}.
\bjtitle{Geophysical Research Letters}
\bvolume{40}(\bissue{6}),
\bfpage{1182}--\blpage{1188}
(\byear{2013})
\doiurl{10.1002/grl.50289}
\end{barticle}
\endbibitem

\bibitem[\protect\citeauthoryear{Drijfhout et~al.}{2011}]{Drijfhout2011}
\begin{barticle}
\bauthor{\bsnm{Drijfhout}, \binits{S.S.}},
\bauthor{\bsnm{Weber}, \binits{S.L.}},
\bauthor{\bsnm{Swaluw}, \binits{E.}}:
\batitle{{The stability of the MOC as diagnosed from model projections for
  pre-industrial, present and future climates}}.
\bjtitle{Climate Dynamics}
\bvolume{37}(\bissue{7-8}),
\bfpage{1575}--\blpage{1586}
(\byear{2011})
\doiurl{10.1007/s00382-010-0930-z}
\end{barticle}
\endbibitem

\bibitem[\protect\citeauthoryear{Mecking et~al.}{2016}]{Mecking2016}
\begin{barticle}
\bauthor{\bsnm{Mecking}, \binits{J.}},
\bauthor{\bsnm{Drijfhout}, \binits{S.S.}},
\bauthor{\bsnm{Jackson}, \binits{L.C.}},
\bauthor{\bsnm{Graham}, \binits{T.}}:
\batitle{Stable amoc off state in an eddy-permitting coupled climate model}.
\bjtitle{Climate Dynamics}
\bvolume{47},
\bfpage{2455}--\blpage{2470}
(\byear{2016})
\end{barticle}
\endbibitem

\bibitem[\protect\citeauthoryear{Bonan et~al.}{2024}]{Bonan2024}
\begin{botherref}
\oauthor{\bsnm{Bonan}, \binits{D.}},
\oauthor{\bsnm{Thompson}, \binits{A.}},
\oauthor{\bsnm{Schneider}, \binits{T.}},
\oauthor{\bsnm{Zanna}, \binits{L.}},
\oauthor{\bsnm{Armour}, \binits{K.}},
\oauthor{\bsnm{Sun}, \binits{S.}}:
Constraints imply limited future weakening of atlantic meridional overturning
  circulation.
Preprint, https://doi.org/10.21203/rs.3.rs-4456168/v1
(2024)
\end{botherref}
\endbibitem

\bibitem[\protect\citeauthoryear{Dijkstra and van Westen}{2024}]{Dijkstra2024b}
\begin{botherref}
\oauthor{\bsnm{Dijkstra}, \binits{H.A.}},
\oauthor{\bsnm{Westen}, \binits{R.M.}}:
{The Effect of Indian Ocean Surface Freshwater Flux Biases On the Multi-Stable
  Regime of the AMOC}.
Tellus A: Dynamic Meteorology and Oceanography
\textbf{76}(1)
(2024)
\end{botherref}
\endbibitem

\bibitem[\protect\citeauthoryear{Lohmann et~al.}{2024}]{Lohmann2024}
\begin{barticle}
\bauthor{\bsnm{Lohmann}, \binits{J.}},
\bauthor{\bsnm{Dijkstra}, \binits{H.A.}},
\bauthor{\bsnm{Jochum}, \binits{M.}},
\bauthor{\bsnm{Lucarini}, \binits{V.}},
\bauthor{\bsnm{Ditlevsen}, \binits{P.D.}}:
\batitle{Multistability and intermediate tipping of the atlantic ocean
  circulation}.
\bjtitle{Science Advances}
\bvolume{10}(\bissue{12}),
\bfpage{4253}
(\byear{2024})
\end{barticle}
\endbibitem

\bibitem[\protect\citeauthoryear{Rahmstorf et~al.}{2005}]{Rahmstorf2005}
\begin{barticle}
\bauthor{\bsnm{Rahmstorf}, \binits{S.}},
\bauthor{\bsnm{Crucifix}, \binits{M.}},
\bauthor{\bsnm{Ganopolski}, \binits{A.}},
\bauthor{\bsnm{Goosse}, \binits{H.}},
\bauthor{\bsnm{Kamenkovich}, \binits{I.}},
\bauthor{\bsnm{Knutti}, \binits{R.}},
\bauthor{\bsnm{Lohmann}, \binits{G.}},
\bauthor{\bsnm{March}, \binits{R.}},
\bauthor{\bsnm{Mysak}, \binits{L.}},
\bauthor{\bsnm{Wang}, \binits{Z.}},
\bauthor{\bsnm{Weaver}, \binits{A.J.}}:
\batitle{{Thermohaline circulation hysteresis: a model intercomparison}}.
\bjtitle{Geophysical Research Letters}
\bvolume{L23605},
\bfpage{1}--\blpage{5}
(\byear{2005})
\end{barticle}
\endbibitem

\bibitem[\protect\citeauthoryear{Drijfhout et~al.}{2015}]{Drijfhout2015}
\begin{barticle}
\bauthor{\bsnm{Drijfhout}, \binits{S.}},
\bauthor{\bsnm{Bathiany}, \binits{S.}},
\bauthor{\bsnm{Beaulieu}, \binits{C.}},
\bauthor{\bsnm{Brovkin}, \binits{V.}},
\bauthor{\bsnm{Claussen}, \binits{M.}},
\bauthor{\bsnm{Huntingford}, \binits{C.}},
\bauthor{\bsnm{Scheffer}, \binits{M.}},
\bauthor{\bsnm{Sgubin}, \binits{G.}},
\bauthor{\bsnm{Swingedouw}, \binits{D.}}:
\batitle{Catalogue of abrupt shifts in intergovernmental panel on climate
  change climate models}.
\bjtitle{Proceedings of the National Academy of Sciences}
\bvolume{112}(\bissue{43}),
\bfpage{5777}--\blpage{5786}
(\byear{2015})
\end{barticle}
\endbibitem

\bibitem[\protect\citeauthoryear{Sgubin et~al.}{2017}]{Sgubin2017}
\begin{barticle}
\bauthor{\bsnm{Sgubin}, \binits{G.}},
\bauthor{\bsnm{Swingedouw}, \binits{D.}},
\bauthor{\bsnm{Drijfhout}, \binits{S.}},
\bauthor{\bsnm{Mary}, \binits{Y.}},
\bauthor{\bsnm{Bennabi}, \binits{A.}}:
\batitle{Abrupt cooling over the north atlantic in modern climate models}.
\bjtitle{Nature Communications}
\bvolume{8}(\bissue{1}),
\bfpage{14375}
(\byear{2017})
\end{barticle}
\endbibitem

\bibitem[\protect\citeauthoryear{Chang et~al.}{2020}]{Chang2020}
\begin{barticle}
\bauthor{\bsnm{Chang}, \binits{P.}},
\bauthor{\bsnm{Zhang}, \binits{S.}},
\bauthor{\bsnm{Danabasoglu}, \binits{G.}},
\bauthor{\bsnm{Yeager}, \binits{S.G.}},
\bauthor{\bsnm{Fu}, \binits{H.}},
\bauthor{\bsnm{Wang}, \binits{H.}},
\bauthor{\bsnm{Castruccio}, \binits{F.S.}},
\bauthor{\bsnm{Chen}, \binits{Y.}},
\bauthor{\bsnm{Edwards}, \binits{J.}},
\bauthor{\bsnm{Fu}, \binits{D.}}, \betal:
\batitle{{An unprecedented set of high-resolution earth system simulations for
  understanding multiscale interactions in climate variability and change}}.
\bjtitle{Journal of Advances in Modeling Earth Systems}
\bvolume{12}(\bissue{12}),
\bfpage{2020}--\blpage{002298}
(\byear{2020})
\end{barticle}
\endbibitem

\bibitem[\protect\citeauthoryear{Gouretski and
  Reseghetti}{2010}]{Gouretski2010}
\begin{barticle}
\bauthor{\bsnm{Gouretski}, \binits{V.}},
\bauthor{\bsnm{Reseghetti}, \binits{F.}}:
\batitle{On depth and temperature biases in bathythermograph data: Development
  of a new correction scheme based on analysis of a global ocean database}.
\bjtitle{Deep Sea Research Part I: Oceanographic Research Papers}
\bvolume{57}(\bissue{6}),
\bfpage{812}--\blpage{833}
(\byear{2010})
\end{barticle}
\endbibitem

\bibitem[\protect\citeauthoryear{Jansen et~al.}{2018}]{Jansen2018}
\begin{barticle}
\bauthor{\bsnm{Jansen}, \binits{M.F.}},
\bauthor{\bsnm{Nadeau}, \binits{L.-P.}},
\bauthor{\bsnm{Merlis}, \binits{T.M.}}:
\batitle{Transient versus equilibrium response of the ocean’s overturning
  circulation to warming}.
\bjtitle{Journal of Climate}
\bvolume{31}(\bissue{13}),
\bfpage{5147}--\blpage{5163}
(\byear{2018})
\end{barticle}
\endbibitem

\bibitem[\protect\citeauthoryear{McDougall and Barker}{2011}]{Mcdougall2011}
\begin{barticle}
\bauthor{\bsnm{McDougall}, \binits{T.J.}},
\bauthor{\bsnm{Barker}, \binits{P.M.}}:
\batitle{{Getting started with TEOS-10 and the Gibbs Seawater (GSW)
  oceanographic toolbox}}.
\bjtitle{Scor/iapso WG}
\bvolume{127}(\bissue{532}),
\bfpage{1}--\blpage{28}
(\byear{2011})
\end{barticle}
\endbibitem

\bibitem[\protect\citeauthoryear{Bil{\'o} et~al.}{2022}]{Bilo2022}
\begin{barticle}
\bauthor{\bsnm{Bil{\'o}}, \binits{T.}},
\bauthor{\bsnm{Straneo}, \binits{F.}},
\bauthor{\bsnm{Holte}, \binits{J.}},
\bauthor{\bsnm{Le~Bras}, \binits{I.-A.}}:
\batitle{{Arrival of new great salinity anomaly weakens convection in the
  Irminger Sea}}.
\bjtitle{Geophysical Research Letters}
\bvolume{49}(\bissue{11}),
\bfpage{2022}--\blpage{098857}
(\byear{2022})
\end{barticle}
\endbibitem

\bibitem[\protect\citeauthoryear{Arum{\'\i}-Planas
  et~al.}{2024}]{ArumiPlanas2024}
\begin{barticle}
\bauthor{\bsnm{Arum{\'\i}-Planas}, \binits{C.}},
\bauthor{\bsnm{Dong}, \binits{S.}},
\bauthor{\bsnm{Perez}, \binits{R.}},
\bauthor{\bsnm{Harrison}, \binits{M.J.}},
\bauthor{\bsnm{Farneti}, \binits{R.}},
\bauthor{\bsnm{Hern{\'a}ndez-Guerra}, \binits{A.}}:
\batitle{{A multi-data set analysis of the freshwater transport by the Atlantic
  meridional overturning circulation at nominally 34.5 S}}.
\bjtitle{Journal of Geophysical Research: Oceans}
\bvolume{129}(\bissue{6}),
\bfpage{2023}--\blpage{020558}
(\byear{2024})
\end{barticle}
\endbibitem

\end{thebibliography}
\end{document}